\documentclass[twoside,english,10pt,journal,a4paper,doublecolumn]{IEEEtran}
\usepackage[T1]{fontenc}
\usepackage[latin9]{inputenc}
\usepackage{textcomp}
\usepackage{amsmath}
\usepackage{amssymb}
\usepackage{graphicx}

\makeatletter

\providecommand{\tabularnewline}{\\}

\usepackage{changebar}

\@ifundefined{showcaptionsetup}{}{%
 \PassOptionsToPackage{caption=false}{subfig}}
\usepackage{subfig}
\makeatother

\begin{document}

\title{Achieving Maximum Energy-Efficiency in Multi-Relay OFDMA Cellular
Networks:\\
A Fractional Programming Approach }

\author{Kent~Tsz~Kan~Cheung, Shaoshi~Yang, and Lajos~Hanzo,~\IEEEmembership{Fellow,~IEEE}\thanks{This research has been funded by the Industrial Companies who are Members of the Mobile VCE, with additional financial support from the UK Government's Engineering \& Physical Sciences Research Council (EPSRC). The financial support of the China Scholarship Council (CSC), of the Research Councils UK (RCUK) under the India-UK Advanced Technology Center (IU-ATC), of the EU under the auspices of the Concerto project, and of the European Research Council's Senior Research Fellow Grant is also gratefully acknowledged.

K. T. K. Cheung is with the School of Electronics and Computer Science, University of Southampton, Southampton, SO17 1BJ, UK (e-mail: ktkc106@ecs.soton.ac.uk).

S. Yang is with the School of Electronics and Computer Science, University of Southampton, Southampton, SO17 1BJ, UK. He is also with the School of Information and Communication Engineering, Beijing University of Posts and Telecommunications, Beijing, 100876, China (e-mail: sy7g09@ecs.soton.ac.uk).

L. Hanzo is with the School of Electronics and Computer Science, University of Southampton, Southampton, SO17 1BJ, UK (e-mail: lh@ecs.soton.ac.uk). }}

\markboth{Published in IEEE Transactions on Communications, vol. 61, no. 7, pp. 2746-2757, Jul. 2013}%
{Shell \MakeLowercase{\textit{et al.}}: Bare Demo of IEEEtran.cls
for Journals}

\maketitle
\vspace{-2cm}

\begin{abstract}
In this paper, the joint power and subcarrier allocation problem is
solved in the context of maximizing the energy-efficiency~(EE) of
a multi-user, multi-relay orthogonal frequency division multiple access~(OFDMA)
cellular network, where the objective function is formulated as the
ratio of the spectral-efficiency~(SE) over the total power dissipation.
It is proven that the fractional programming problem considered is
quasi-concave so that Dinkelbach's method may be employed for finding
the optimal solution at a low complexity. This method solves the above-mentioned
master problem by solving a series of parameterized concave secondary
problems. These secondary problems are solved using a dual decomposition
approach, where each secondary problem is further decomposed into
a number of similar subproblems. The impact of various system parameters
on the attainable EE and SE of the system employing both EE maximization~(EEM)
and SE maximization~(SEM) algorithms is characterized. In particular,
it is observed that increasing the number of relays for a range of
cell sizes, although marginally increases the attainable SE, reduces
the EE significantly. It is noted that the highest SE and EE are achieved,
when the relays are placed closer to the BS to take advantage of the
resultant line-of-sight link. Furthermore, increasing both the number
of available subcarriers and the number of active user equipment~(UE)
increases both the EE and the total SE of the system as a benefit
of the increased frequency and multi-user diversity, respectively.
Finally, it is demonstrated that as expected, increasing the available
power tends to improve the SE, when using the SEM algorithm. By contrast,
given a sufficiently high available power, the EEM algorithm attains
the maximum achievable EE and a suboptimal SE.\end{abstract}

\section{Introduction\label{sec:Introduction}}

\IEEEPARstart{E}{nergy}-efficiency~(EE) is becoming of great concern
in the telecommunications community owing to the rapidly increasing
data rate requirements, increasing energy prices, and societal as
well as political pressures on mobile phone operators to reduce their
'carbon footprint'~\cite{Correia2010}. This has led to several joint
academic and industrial research efforts dedicated to developing novel
energy-saving techniques, such as the 'green radio' project~\cite{Han2011},
the GreenTouch alliance~\cite{Collins2010}, and the energy aware
radio and network technologies~(EARTH) project~\cite{EARTH}. Substantial
research efforts have also been dedicated to the next-generation wireless
networks, such as the third generation partnership project's~(3GPP)
long term evolution-advanced~(LTE-A) and IEEE 802.16 worldwide interoperability
for microwave access~(WiMAX)~\cite{Yang2009} standards, which may
rely on relaying between the central base station~(BS) and the user
equipment~(UE). As a benefit of reduced transmission distances, either
the quality of the communication is maintained at reduced power requirements,
or the transmission integrity is improved at the same power consumption.
This allows the need for expensive deployment and maintenance of additional
BSs to be circumvented. The two most popular relaying techniques are
the amplify-and-forward~(AF) and the decode-and-forward~(DF) schemes~\cite{Laneman2004}.
The AF regime is less complex than DF, since the relay node~(RN)
needs only to receive and linearly amplify the source\textquoteright{}s
transmissions, before forwarding it to the destination. The effects
of scheduling and frequency reuse in the context of the above-mentioned
networks was studied in~\cite{Oyman2010}.

Both LTE-A and WiMAX employ the orthogonal frequency division multiple
access~(OFDMA) technique. In OFDMA, the whole channel's bandwidth
is divided into multiple subcarriers, where subsets of subcarriers
may be allocated for transmission to different users~\cite{Hanzo2010}.
In OFDMA, the system attains two types of diversity, which may be
jointly exploited for improving the achievable sum-rate~(SR) of the
system. Firstly, multi-user diversity is attained with the aid of
appropriate user mapping, because when the channel from the BS to
a specific UE is undergoing severe fading on a particular subcarrier,
then this subcarrier may be assigned for transmission to another user,
whose channel might be more friendly. On the other hand, activating
only those subcarriers that are suitable for high-quality transmission
to a particular UE leads to frequency diversity. These philosophies
underpin several contributions in the literature, where the goal is
to assign the available resources, for example power and subcarriers,
so that a system-wide metric is maximized. These methods belong to
the family of resource allocation policies and typically aim for solving
one of two problems: either the spectral-efficiency~(SE)%
\footnote{For a given system bandwidth, the SR maximization and SE maximization~(SEM)
solutions are identical. To avoid any additional abbreviations, they
are both henceforth referred to as SEM.%
}~\cite{Li2006,Ng2007,Ng2010} of the system is maximized while a
maximum power constraint is enforced, or the power consumption is
minimized under a minimum total system throughput or individual UE
rate constraint%
\footnote{In the latter case, the minimum rate constraint may be viewed as ensuring
fairness among the users, since each user achieves at least a minimum
rate.%
}~\cite{Wong1999,Piazzo1999,Xiao2008,Piazzo2011,Joung2012}.

An example of the SE Maximization~(SEM) problem was considered in~\cite{Ng2010},
where the authors formulate the optimization problem for the downlink~(DL)
of an AF relaying-aided OFDMA cellular network, and their goal was
to optimize the power and subcarrier allocation so that the SE of
the system was maximized under maximum outage probability and total
power constraints. In the class of power minimization problems, an
example is the often-cited work by Wong \emph{et. al}~\cite{Wong1999},
where a heuristic bit allocation algorithm was conceived for a multi-user
OFDMA system with the aim of minimizing the power consumption under
a minimum individual user rate constraint. With a similar goal, Piazzo~\cite{Piazzo1999}
developed a sub-optimal bit allocation algorithm for an orthogonal
frequency division multiplexing~(OFDM) system. This work was later
extended to provide the optimal bit allocation in~\cite{Piazzo2011}.

However, the SEM and the power minimization problems do not directly
consider an EE objective function~(OF), and in general they do not
deliver the EE maximization~(EEM) solution. In recent years, research
into resource allocation using an EE OF has become increasingly popular.
In reality, EEM may be viewed as an example of multi-objective optimization,
since typically the goal is to maximize the SE achieved, whilst concurrently
minimizing the power consumption required. From this perspective,~\cite{Devarajan2012}
derives an aggregate OF, which consists of a weighted sum of the SR
achieved and the power dissipated. However, selecting appropriate
weights for the two OFs is not straightforward, and different combinations
of weights can lead to very different results. Another example is
given in~\cite{Yu2011}, where the EEM problem is considered in a
multi-relay network employing the AF protocol. However, both~\cite{Devarajan2012,Yu2011}
only optimize the user selection and power allocation without considering
the subcarrier allocation in the network. Another formulation, demonstrated
in~\cite{Miao2010,Miao2012}, considers power and subcarrier allocation
in an OFDMA cellular network, but without a maximum total power constraint
and without relaying. The authors of~\cite{Ng2012} formulate the
EEM problem in a OFDMA cellular network under a maximum total power
constraint, however relaying is not considered.

In light of the above discussions, this work focuses on a solution
method for the EEM problem in a multi-relay, multi-user OFDMA cellular
network, which jointly considers both power and subcarrier allocation
as well as a maximum total power constraint. The contributions of
this paper is summarized as follows:
\begin{itemize}
\item The EEM problem in the context of a multi-relay, multi-user OFDMA
cellular network, in which both direct and relayed transmissions are
employed, is formulated as a fractional programming problem, which
jointly considers both the power and subcarrier allocation. In contrast
to previous contributions such as~\cite{Oyman2010}, the aim is for
finding the optimal power and subcarrier allocations within a network
context. Furthermore, in contrast to~\cite{Wong1999,Piazzo1999,Li2006,Ng2007,Xiao2008,Ng2010,Piazzo2011,Yu2011,Devarajan2012},
the focus is placed on an EE OF. It is demonstrated that in its original
form the problem is a mixed-integer non-linear programming problem~(MINLP)~\cite{Bertsekas1999},
which is challenging to solve. In order to make the problem more tractable,
both a variable transformation and a relaxation of the integer variables
is introduced.
\item It is proven that the relaxed problem is quasi-concave and consequently
Dinkelbach's method~\cite{Dinkelbach1967} may be employed for obtaining
the optimal solution by solving a sequence of parameterized secondary
problems. Each of these are solved using the dual decomposition approach
of~\cite{Palomar2006}. It is demonstrated that the EEM algorithm
reaches the optimal solution within a low number of iterations and
reaches the optimal solution obtained via an exhaustive search. Thus
the original problem is solved at a low complexity.
\item Comparisons are made between two multi-relay resource allocation problems,
namely one that solves the EEM problem and another that considers
SEM. As an example, it is shown that when the maximum affordable power
is lower than a given threshold, the two problems have the same solutions.
However, as the maximum affordable power is increased, the SEM algorithm
attempts to achieve a higher SE at the cost of a lower EE, while given
the total power, the EEM algorithm reaches the upper limit of the
maximum achievable SE for the sake of maintaining the maximum EE.
\item Since the system model is generalized, the EEM and SEM algorithms
may be employed for gaining insights into network design, when the
aim is for maximizing either the EE or SE. To that end, a comprehensive
range of results are presented, which demonstrate both the effect
of increasing the number of available subcarriers and UEs in the system,
as well as quantifying the impact of increasing the number of RNs
in the system and its relation not only to the cell radius, but also
to the relays' positions. The algorithm may be used for characterizing
the effects of many other system design choices on the maximum SE
and EE.
\end{itemize}
The rest of this paper is organized as follows. In Section~\ref{sec:System-Model},
the multi-user, multi-relay OFDMA cellular network model is described,
which is followed by a formulation of the optimization problem in
Section~\ref{sec:Problem-Formulation}. Upon invoking a transformation
of variables and a relaxation of the integer variables, it is proven
that the OF is quasi-concave. The combined solution algorithm of Dinkelbach's
method~\cite{Dinkelbach1967} and dual decomposition~\cite{Palomar2006}
is outlined in Section~\ref{sec:Dinkelbach}. The performance of
the EEM algorithm is demonstrated in Section~\ref{sec:Results-and-Discussions},
which includes results obtained when the EEM and SEM algorithms are
employed for characterizing the effect of different system design
choices on the achievable SE and EE. Lastly, conclusions are given
in Section~\ref{sec:Conclusions}, where future work ideas are also
listed.

\section{System Model\label{sec:System-Model}}

\begin{figure}
\centering{}\includegraphics[scale=0.5]{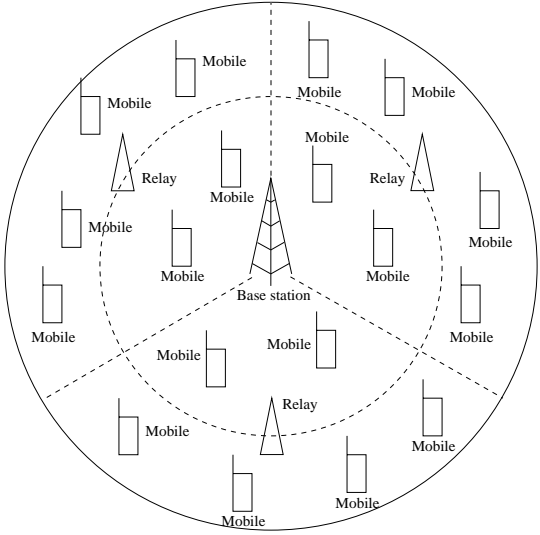}\caption{An example of a cellular network with $M=3$ RNs and $K=18$ UEs.}
\label{fig:cellular}
\end{figure}
Consider an OFDMA DL cellular system relying on a single BS, $M$
fixed RNs and $K$ uniformly-distributed UEs, as shown in Fig.~\ref{fig:cellular}%
\footnote{Although it is more realistic to consider a multi-cell system, which
would lead to inter-cell interference, our system model assumes that
intelligent interference coordination or mitigation techniques are
employed such that the level of inter-cell interference is negligible~\cite{Gesbert2010}.%
}. The cell is divided into $M$ sectors, where each sector is served
by one of the fixed RNs. Naturally, the path-loss is a major factor
in determining the receiver's signal-to-noise ratios~(SNRs) at the
UEs, and thus has a substantial effect on the EE. Therefore, in order
to minimize the RN-to-UE pathloss, all the UEs in a specific sector
are only supported by that sector's RN, and therefore relay selection
is implicitly accomplished. Although the model may be readily extended
to include relay selection, for the sake of mathematical tractability,
it is not included in this work. The model accounts for both the AF
relayed link as well as for the direct link between the BS and UEs,
while the variables related to these two communication protocols are
distinguishable by the superscripts $A$ and $D$, respectively. When
defining links, the subscript $0$ is used for indicating the BS,
whilst $\mathcal{M}(k)\in\{1,\cdots,M\}$ indicates the RN selected
for assisting the DL-transmissions to user $k$. The proportion of
the BS-to-RN distance to the cell radius is denoted by $D_{r}$, while
the total available instantaneous transmission power of the network
is $P_{max}$. Although it is more realistic to consider a system
with separate power constraints for each transmitting entity, for
simplicity, a certain total power constraint is considered%
\footnote{Additionally, it was empirically shown the dual decomposition approach
only obtains a local optimum when separate BS and RN power constraints
are imposed.%
}. The results obtained provide insights into holistic system design
by granting a higher grade of freedom in terms of sharing the power
among the transmitting entities, and thus attaining a higher performance.

Using the direct transmission protocol, the receiver's SNR at UE $k$
on subcarrier $n$ may be expressed as $\Gamma_{k}^{D,n}(\mathcal{P})$,
whereas when using the AF relaying protocol, the receiver's SNR at
UE $k$ on subcarrier $n$ may be expressed as~\cite{Laneman2004}
\begin{equation}
\Gamma_{k}^{A,n}(\mathcal{P})=\frac{\gamma_{0,\mathcal{M}(k)}^{A,n}\gamma_{\mathcal{M}(k),k}^{A,n}}{\left(\gamma_{0,\mathcal{M}(k)}^{A,n}+\gamma_{\mathcal{M}(k),k}^{A,n}+1\right)},\label{eq:snr}
\end{equation}
where $\gamma_{a,b}^{X,n}=P_{a,b}^{X,n}G_{a,b}^{n}/\Delta\gamma N_{0}W$
is the SNR at receiver $b\in\{1,\cdots,M,1,\cdots,K\}$ on subcarrier
$n\in\{1,\cdots,N\}$, and $P_{a,b}^{X,n}$ is allocated to transmitter
$a\in\{0,\cdots,M\}$ using protocol $X\in\{D,A\}$ for transmission
to receiver $b$. Furthermore, $G_{a,b}^{n}$ represents the channel's
attenuation between transmitter $a$ and receiver $b$ on subcarrier
$n$, which is assumed to be known at the BS for all links. The channel's
attenuation is modeled by the path-loss and the Rayleigh fading between
the transmitter and receiver. Furthermore, $N_{0}$ is the additive
white Gaussian noise~(AWGN) variance and $W$ is the bandwidth of
a single subcarrier. Still referring to~(\ref{eq:snr}), $\Delta\gamma$
is the SNR gap at the system's bit error ratio~(BER) target between
the SNR required at the discrete-input continuous-output memoryless
channel~(DCMC) capacity and the actual SNR required the modulation
and coding schemes of the practical physical layer transceivers employed.
For example, making the simplifying assumption that idealized transceivers
operating exactly at the DCMC capacity are employed, then $\Delta\gamma=0\mbox{ dB}$.
Although it is not possible to operate exactly at the DCMC channel
capacity, several physical layer transceiver designs exist that operate
arbitrarily close to it~\cite{Hanzo2009}. Additionally, the power
allocation policy of the system is denoted by $\mathcal{P}$, which
determines the values of $P_{a,b}^{X,n}$.

Assuming sufficiently high receiver's SNR values, the following approximation
can be made
\begin{equation}
\Gamma_{k}^{A,n}(\mathcal{P})\approx\frac{P_{0,\mathcal{M}(k)}^{A,n}G_{0,\mathcal{M}(k)}^{n}P_{\mathcal{M}(k),k}^{A,n}G_{\mathcal{M}(k),k}^{n}}{\Delta\gamma N_{0}W\left(P_{0,\mathcal{M}(k)}^{A,n}G_{0,\mathcal{M}(k)}^{n}+P_{\mathcal{M}(k),k}^{A,n}G_{\mathcal{M}(k),k}^{n}\right)},
\end{equation}
which is valid%
\footnote{It is plausible that in next-generation systems, through the combination
of multi-user and frequency diversity, this assumption holds true
when an intelligent scheduler is employed~\cite{Oyman2010}.%
} for $P_{0,\mathcal{M}(k)}^{A,n}G_{0,\mathcal{M}(k)}^{n}+P_{\mathcal{M}(k),k}^{A,n}G_{\mathcal{M}(k),k}^{n}\gg\Delta\gamma N_{0}W$.
The SE of an AF link to UE $k$ on subcarrier $n$ is then given by
\begin{equation}
R_{k}^{A,n}(\mathcal{P})=\frac{1}{2}\mbox{log}_{2}\left(1+\Gamma_{k}^{A,n}\right)\mbox{ }\mbox{[bits/s/Hz]},
\end{equation}
where the factor of $\frac{1}{2}$ accounts for the fact that two
time slots are required for the two-hop AF transmission. The SE of
a direct link to UE $k$ on subcarrier $n$ is similarly given by
\begin{equation}
R_{k}^{D,n}(\mathcal{P})=\mbox{log}_{2}\left(1+\Gamma_{k}^{D,n}\right)\mbox{ }\mbox{[bits/s/Hz]}.
\end{equation}

The subcarrier indicator variable $s_{k}^{X,n}\in\{0,1\}$ is now
introduced, which denotes the allocation of subcarrier $n$ for transmission
to user $k$ using protocol $X$ for $s_{k}^{X,n}=1$, and $s_{k}^{X,n}=0$
otherwise. The weighted average SE of the system is calculated as 

\begin{eqnarray}
R_{T}(\mathcal{P},\mathcal{S}) & = & \frac{1}{N}\sum_{k=1}^{K}\omega_{k}\sum_{n=1}^{N}s_{k}^{D,n}\mbox{log}_{2}\left(1+\Gamma_{k}^{D,n}\right)\nonumber \\
 &  & +\frac{s_{k}^{A,n}}{2}\mbox{log}_{2}\left(1+\Gamma_{k}^{A,n}\right)\mbox{\mbox{ }[bits/s/Hz]},
\end{eqnarray}
where $\mathcal{S}$ denotes the subcarrier allocation policy of the
system, which determines the values of the subcarrier indicator variable
$s_{k}^{X,n}$. The weighting factor $\omega_{k}$ may be varied for
ensuring fairness amongst users. However, since ensuring fairness
is not the focus of this work, $\omega_{k}=1$, $\forall k$ is assumed
then the effect of $\omega_{k}$ may be ignored.

In order to compute the energy used in these transmissions, a model
similar to~\cite{Arnold2010} is adopted and the total power consumption
of the system is assumed be governed by a constant term and a term
that varies with the transmission powers, which may be written as~(\ref{eq:P_T}).

\begin{figure*}[tbp]
\begin{equation}
P_{T}(\mathcal{P},\mathcal{S})=\left(P_{C}^{(B)}+M\cdot P_{C}^{(R)}\right)+\sum_{k=1}^{K}\sum_{n=1}^{N}s_{k}^{D,n}\xi^{(B)}P_{0,k}^{D,n}+\frac{1}{2}s_{k}^{A,n}\cdot\left(\xi^{(B)}P_{0,\mathcal{M}(k)}^{A,n}+\xi^{(R)}P_{\mathcal{M}(k),k}^{A,n}\right)\mbox{ }\mbox{[Watts]}\label{eq:P_T}
\end{equation}
\hrulefill
\end{figure*}

Here, $P_{C}^{(B)}$ and $P_{C}^{(R)}$ represent the fixed power
consumption of each BS and each RN, respectively, while $\xi^{(B)}>1$
and $\xi^{(R)}>1$ denote the reciprocal of the drain efficiencies
of the power amplifiers employed at the BS and the RNs, respectively.
For example, an amplifier having a $25\%$ drain efficiency would
have $\xi=\frac{1}{0.25}=4$. 

Finally, the average EE metric of the system is expressed as
\begin{equation}
\eta_{E}(\mathcal{P},\mathcal{S})=\frac{R_{T}(\mathcal{P},\mathcal{S})}{P_{T}(\mathcal{P},\mathcal{S})}\mbox{ [bits/Joule/Hz]}.\label{eq:eeff}
\end{equation}

\section{Problem Formulation\label{sec:Problem-Formulation}}

The aim of this work is to maximize the energy efficiency metric of~(\ref{eq:eeff})
subject to a maximum total instantaneous transmit power constraint.
In its current form,~(\ref{eq:eeff}) is dependent on $3KN$ continuous
power variables $P_{0,k}^{D,n}$, $P_{0,\mathcal{M}(k)}^{A,n}$ and
$P_{\mathcal{M}(k),k}^{A,n}$, $\forall k,n$, and $2KN$ binary subcarrier
indicator variables $s_{k}^{D,n}$ and $s_{k}^{A,n}$, $\forall k,n$.
Thus, it may be regarded as a MINLP problem~\cite{Bertsekas1999},
and can be solved using the branch-and-bound method of~\cite{Boyd2007a}.
However, the computational effort required for branch-and-bound techniques
typically increases exponentially with the problem size. Therefore,
a simpler solution is derived by relaxing the binary constraint imposed
on the subcarrier indicator variables, $s_{k}^{D,n}$ and $s_{k}^{A,n}$,
so that they may assume continuous values from the interval $[0,1]$,
as demonstrated in~\cite{Wong1999,Wei2006}. Furthermore, the variables
$\widetilde{P}_{0,k}^{D,n}=P_{0,k}^{D,n}s_{k}^{D,n}$, $\widetilde{P}_{0,\mathcal{M}(k)}^{A,n}=P_{0,\mathcal{M}(k)}^{A,n}s_{k}^{A,n}$
and $\widetilde{P}_{0,\mathcal{M}(k)}^{A,n}=P_{0,\mathcal{M}(k)}^{A,n}s_{k}^{A,n}$
are introduced.

The relaxation of the binary constraints imposed on the variables
$s_{k}^{D,n}$ and $s_{k}^{A,n}$ allows them to assume continuous
values, which leads to a time-sharing subcarrier allocation between
the UEs. Naturally, the original problem is not actually solved. However,
it has been shown that solving the dual of the relaxed problem provides
solutions that are arbitrarily close to the original, non-relaxed
problem, provided that the number of available subcarriers tends to
infinity~\cite{Wei2006}. It has empirically been shown that in some
cases only $8$ subcarriers are required for obtaining close-to-optimal
results~\cite{Seong2006}. It shall be demonstrated in Section~\ref{sec:Results-and-Discussions}
that even for as few as two subcarriers, the solution algorithm employed
in this work approaches the optimal EE achieved by an exhaustive search.

The optimization problem is formulated as shown as follows:

Relaxed Problem (P):
\begin{eqnarray}
\mbox{maximize} &  & \frac{\widetilde{R}_{T}}{\widetilde{P}_{T}}\label{eq:eeff_orig}\\
\mbox{}\nonumber \\
\mbox{ subject to} &  & \sum_{k=1}^{K}\sum_{n=1}^{N}\widetilde{P}_{0,k}^{D,n}+\widetilde{P}_{0,\mathcal{M}(k)}^{A,n}+\widetilde{P}_{\mathcal{M}(k),k}^{A,n}\leq P_{max},\nonumber \\
\label{eq:C1}\\
 &  & s_{k}^{D,n}+s_{k}^{A,n}\leq1\mbox{, }\forall k,n,\label{eq:C2}\\
 &  & \sum_{k=1}^{K}s_{k}^{D,n}+s_{k}^{A,n}\leq1\mbox{, }\forall n,\label{eq:C3}\\
 &  & \widetilde{P}_{0,k}^{D,n}\mbox{, }\widetilde{P}_{0,\mathcal{M}(k)}^{A,n}\mbox{, }\widetilde{P}_{\mathcal{M}(k),k}^{A,n}\in\mathbb{R}_{+}\mbox{, }\forall k,n,\label{eq:C4}\\
 &  & 0\leq s_{k}^{D,n}\mbox{, }s_{k}^{A,n}\leq1\mbox{, }\forall k,n,\label{eq:C5}
\end{eqnarray}
where the objective function is the ratio between~(\ref{eq:RR_T})
and~(\ref{eq:PP_T}).

\begin{figure*}[tbp]
\begin{eqnarray}
\widetilde{R}_{T} & = & \sum_{k=1}^{K}\sum_{n=1}^{N}s_{k}^{D,n}\mbox{log}_{2}\left(1+\frac{\widetilde{P}_{0,k}^{D,n}G_{0,k}^{n}}{s_{k}^{D,n}\Delta\gamma N_{0}W}\right)\nonumber \\
 &  & +\frac{s_{k}^{A,n}}{2}\mbox{log}_{2}\left(1+\frac{\widetilde{P}_{0,\mathcal{M}(k)}^{A,n}G_{0,\mathcal{M}(k)}^{n}\widetilde{P}_{\mathcal{M}(k),k}^{A,n}G_{\mathcal{M}(k),k}^{n}}{s_{k}^{A,n}\Delta\gamma N_{0}W\left(\widetilde{P}_{0,\mathcal{M}(k)}^{A,n}G_{0,\mathcal{M}(k)}^{n}+\widetilde{P}_{\mathcal{M}(k),k}^{A,n}G_{\mathcal{M}(k),k}^{n}\right)}\right)\label{eq:RR_T}
\end{eqnarray}
\begin{equation}
\widetilde{P}_{T}=\left(P_{C}^{(B)}+M\cdot P_{C}^{(R)}\right)+\sum_{k=1}^{K}\sum_{n=1}^{N}\xi^{(B)}\widetilde{P}_{0,k}^{D,n}+\frac{1}{2}\left(\xi^{(B)}\widetilde{P}_{0,\mathcal{M}(k)}^{A,n}+\xi^{(R)}\widetilde{P}_{\mathcal{M}(k),k}^{A,n}\right)\label{eq:PP_T}
\end{equation}
\hrulefill
\end{figure*}

In this formulation, the variables to be optimized are $s_{k}^{D,n}$,
$s_{k}^{A,n}$, $\widetilde{P}_{0,k}^{D,n}$, $\widetilde{P}_{0,\mathcal{M}(k)}^{A,n}$
and $\widetilde{P}_{\mathcal{M}(k),k}^{A,n}$, $\forall k,n$. Physically,
the constraint~(\ref{eq:C1}) ensures that the sum of the power allocated
to variables $\widetilde{P}_{0,k}^{D,n}$, $\widetilde{P}_{0,\mathcal{M}(k)}^{A,n}$
and $\widetilde{P}_{\mathcal{M}(k),k}^{A,n}$ does not exceed the
maximum power budget of the system. Constraint~(\ref{eq:C2}) ensures
that a single transmission protocol, either direct or AF, is chosen
for each user-subcarrier pair. The constraint~(\ref{eq:C3}) guarantees
that each subcarrier is only allocated to at most one user, thus intra-cell
interference is avoided. The constraints~(\ref{eq:C4}) and~(\ref{eq:C5})
describe the feasible region of the optimization variables. The following
is a proof that the OF of problem~(P) is quasi-concave~\cite{Avriel1988}.

\subsection{Proving that the OF in problem (P) is quasi-concave}

A function, $f:\mbox{ }\mathbf{R}^{n}\rightarrow\mathbf{R}$, is quasi-concave
if its domain is convex, and all its superlevel sets are convex, i.e.
if the domain $S_{\alpha}=\{x\in\mathbf{dom}\mbox{ }f\mbox{ }|\mbox{ }f(x)\geq\alpha\}$
is convex for $\alpha\in\mathbf{R}$~\cite{Boyd2004}. For a fractional
function, $g(x)/h(x)$, the inequality $g(x)/h(x)\geq\alpha$ is equivalent
to $[g(x)-\alpha h(x)]\geq0$, assuming $h(x)>0$, $\forall x$. Therefore,
in order to prove that~(\ref{eq:eeff_orig}) is quasi-concave, it
is sufficient to show that the numerator is concave and the denominator
is both affine and positive, whilst the domain is convex. It is plausible
that the denominator is both affine and positive, since it is the
linear combination of multiple nonnegative variables and a positive
constant. The proof that the numerator is concave is as follows.

Firstly, the concavity of $f_{1}\left(\widetilde{P}_{0,\mathcal{M}(k)}^{A,n},\widetilde{P}_{\mathcal{M}(k),k}^{A,n}\right)=\frac{\widetilde{P}_{0,\mathcal{M}(k)}^{A,n}G_{0,\mathcal{M}(k)}^{n}\widetilde{P}_{\mathcal{M}(k),k}^{A,n}G_{\mathcal{M}(k),k}^{n}}{\Delta\gamma N_{0}W\left(\widetilde{P}_{0,\mathcal{M}(k)}^{A,n}G_{0,\mathcal{M}(k)}^{n}+\widetilde{P}_{\mathcal{M}(k),k}^{A,n}G_{\mathcal{M}(k),k}^{n}\right)}$
is proven. This may be accomplished by examining the Hessian matrix
of $f_{1}\left(\widetilde{P}_{0,\mathcal{M}(k)}^{A,n},\widetilde{P}_{\mathcal{M}(k),k}^{A,n}\right)$
with respect to~(w.r.t.) the variables $\widetilde{P}_{0,\mathcal{M}(k)}^{A,n}$
and $\widetilde{P}_{\mathcal{M}(k),k}^{A,n}$~\cite{Boyd2004}. The
Hessian has the eigenvalues $e_{1}=0$ and
\begin{equation}
e_{2}=-\frac{2\left(G_{0,\mathcal{M}(k)}^{n}G_{\mathcal{M}(k),k}^{n}\right)^{2}\left(\widetilde{P}_{0,\mathcal{M}(k)}^{A,n}+\widetilde{P}_{\mathcal{M}(k),k}^{A,n}\right)}{\Delta\gamma N_{0}W\left(\widetilde{P}_{0,\mathcal{M}(k)}^{A,n}G_{0,\mathcal{M}(k)}^{n}+\widetilde{P}_{\mathcal{M}(k),k}^{A,n}G_{\mathcal{M}(k),k}^{n}\right)^{3}},
\end{equation}
which are non-positive, indicating that the Hessian is negative-semidefinite.
This indicates that $f_{1}\left(\widetilde{P}_{0,\mathcal{M}(k)}^{A,n},\widetilde{P}_{\mathcal{M}(k),k}^{A,n}\right)$
is concave w.r.t. the variables $\widetilde{P}_{0,\mathcal{M}(k)}^{A,n}$
and $\widetilde{P}_{\mathcal{M}(k),k}^{A,n}$.

Examination of the composition $f_{2}\left(\widetilde{P}_{0,\mathcal{M}(k)}^{A,n},\widetilde{P}_{\mathcal{M}(k),k}^{A,n}\right)=\log_{2}\left[1+f_{1}\left(\widetilde{P}_{0,\mathcal{M}(k)}^{A,n},\widetilde{P}_{\mathcal{M}(k),k}^{A,n}\right)\right]$
reveals that $f_{2}\left(\widetilde{P}_{0,\mathcal{M}(k)}^{A,n},\widetilde{P}_{\mathcal{M}(k),k}^{A,n}\right)$
is concave, since $\log_{2}(\cdot)$ is concave as well as non-decreasing
and $1+f_{1}\left(\widetilde{P}_{0,\mathcal{M}(k)}^{A,n},\widetilde{P}_{\mathcal{M}(k),k}^{A,n}\right)$
is concave~\cite{Boyd2004}. 

The second term in the summation of~(\ref{eq:RR_T}) may be denoted
by~(\ref{eq:f_3}). \begin{figure*}[tbp]
\begin{equation}
f_{3}\left(\widetilde{P}_{0,\mathcal{M}(k)}^{A,n},\widetilde{P}_{\mathcal{M}(k),k}^{A,n},s_{k}^{A,n}\right)=s_{k,n}^{A,n}\cdot\mbox{log}_{2}\left(1+\frac{\widetilde{P}_{0,\mathcal{M}(k)}^{A,n}G_{0,\mathcal{M}(k)}^{n}\widetilde{P}_{\mathcal{M}(k),k}^{A,n}G_{\mathcal{M}(k),k}^{n}}{s_{k}^{A,n}\Delta\gamma N_{0}W\left(\widetilde{P}_{0,\mathcal{M}(k)}^{A,n}G_{0,\mathcal{M}(k)}^{n}+\widetilde{P}_{\mathcal{M}(k),k}^{A,n}G_{\mathcal{M}(k),k}^{n}\right)}\right).\label{eq:f_3}
\end{equation}
\hrulefill
\end{figure*}This may be obtained using the perspective transformation%
\footnote{The perspective transformation of the function $f(x)$ is given by
$tf(x/t)$.%
} of~\cite{Boyd2004} yielding 
\begin{equation}
f_{3}\left(\widetilde{P}_{0,\mathcal{M}(k)}^{A,n},\widetilde{P}_{\mathcal{M}(k),k}^{A,n},s_{k}^{A,n}\right)=s_{k}^{A,n}\cdot f_{2}\left(\frac{\widetilde{P}_{0,\mathcal{M}(k)}^{A,n}}{s_{k}^{A,n}},\frac{\widetilde{P}_{\mathcal{M}(k),k}^{A,n}}{s_{k}^{A,n}}\right),
\end{equation}
which preserves concavity. Using similar arguments, $s_{k}^{D,n}\mbox{log}_{2}\left(1+\frac{\widetilde{P}_{0,k}^{D,n}G_{0,k}^{n}}{s_{k}^{D,n}\Delta\gamma N_{0}W}\right)$
is proven to be concave w.r.t. the variables $s_{k}^{D,n}$ and $\widetilde{P}_{0,k}^{D,n}$.

Finally, the numerator is shown to be concave w.r.t the variables
$s_{k}^{A,n}$, $s_{k}^{D,n}$, $\widetilde{P}_{0,k}^{D,n}$, $\widetilde{P}_{0,k}^{A,n}$,
$\widetilde{P}_{0,\mathcal{M}(k)}^{A,n}$ and $\widetilde{P}_{\mathcal{M}(k),k}^{A,n}$,
$\forall k,n$, since it is the non-negative sum of multiple concave
functions. Thus, the OF in problem~(P) has a numerator that is concave,
while its denominator is affine. Hence, the OF of problem~(P) is
quasi-concave.

\subsection{Problem solution methods}

Quasi-concavity may be viewed as a type of generalized concavity,
since it can describe discontinuous functions as well as functions
that have multiple stationary points. This means that a local maximum
is not guaranteed to be a global maximum and thus standard convex
optimization techniques, such as interior-point or ellipsoid methods,
cannot be readily applied for finding the optimal solution~\cite{Boyd2004}.
However, a quasi-concave function has convex superlevel sets, hence
the bisection method~\cite{Yu2011} may be used for iteratively closing
the gap between an upper and lower bound solution, until the difference
between the two becomes lower than a predefined tolerance. The drawback
of this method is that there is no exact method of finding the initial
upper as well as lower bounds. Additionally, a convex feasibility
problem~\cite{Boyd2004} must be solved in each iteration, which
may become computationally undesirable. In light of these discussions,
the method detailed in~\cite{Dinkelbach1967} is employed, which
allows the quasi-concave problem to be solved as a sequence of parameterized
concave programming problems. For clarity, the algorithm is summarized
in Fig.~\ref{fig:summary}, which is discussed in the following section.

\section{Dinkelbach's method for solving problem (P)\label{sec:Dinkelbach}}

\subsection{Introduction to Dinkelbach's method\label{sub:Dinkelbach_outer}}

\begin{table}
\setlength{\arrayrulewidth}{1pt}

\centering{}\caption{Dinkelbach's method for energy efficiency maximization.}
\label{algor:dinkelbach}%
\begin{tabular}{rl}
\hline 
\multicolumn{2}{l}{\textbf{Algorithm 1} Dinkelbach's method for energy efficiency maximization}\tabularnewline
\hline 
Input: & $I_{outer}^{D}$~(maximum number of iterations)\tabularnewline
 & $\epsilon_{outer}^{D}>0$~(convergence tolerance)\tabularnewline
 & \tabularnewline
1: & $q_{0}\leftarrow0$\tabularnewline
2: & $i\leftarrow0$\tabularnewline
3: & \textbf{do while} $q_{i}-q_{i-1}>\epsilon_{outer}^{D}$ \textbf{and}
$i<I_{outer}^{D}$ \tabularnewline
4: & \quad{}$i\leftarrow i+1$\tabularnewline
5: & \quad{}Solve $\underset{\mathcal{P},\mathcal{S}}{\mbox{max.}}R_{T}(\mathcal{P},\mathcal{S})-q_{i-1}P_{T}(\mathcal{P},\mathcal{S})$
to obtain the\tabularnewline
 & \quad{}optimal solution $\mathcal{P}_{i}^{*}$ and $\mathcal{S}_{i}^{*}$
(inner loop)\tabularnewline
6: & \quad{}$q_{i}\leftarrow\mbox{ }R_{T}(\mathcal{P}_{i}^{*},\mathcal{S}_{i}^{*})/P_{T}(\mathcal{P}_{i}^{*},\mathcal{S}_{i}^{*})$\tabularnewline
7: & \textbf{end do}\tabularnewline
8: & \textbf{return}\tabularnewline
\hline 
\end{tabular}\vspace{-0.6cm}
\end{table}
Dinkelbach's method~\cite{Dinkelbach1967,Avriel1988} is an iterative
algorithm that can be used for solving a quasi-concave problem in
a parameterized concave form. The algorithm is summarized in Table~\ref{algor:dinkelbach}.
The concave form of the fractional program~(P) is formed by denoting
the OF value as $q$ so that a subtractive form of the OF may be written
as $F(q)=R_{T}(\mathcal{P},\mathcal{S})-qP_{T}(\mathcal{P},\mathcal{S})$,
which is concave. Since the parameter $q$ now acts as a negative
weight on the total power consumption of system, it may be intuitively
viewed as the 'price' of the system's power consumption. At the optimal
OF value of $q^{*}$, the following holds true 
\begin{equation}
\underset{\mathcal{P},\mathcal{S}}{\mbox{max.}}\left\{ F(q^{*})\right\} =\underset{\mathcal{P},\mathcal{S}}{\mbox{max.}}\left\{ R_{T}(\mathcal{P},\mathcal{S})-q^{*}P_{T}(\mathcal{P},\mathcal{S})\right\} =0.
\end{equation}
Explicitly, the solution of $F(q^{*})$ is equivalent to the solution
of the fractional problem~(P). Dinkelbach~\cite{Dinkelbach1967}
proposed an iterative method to find increasing $q$ values, which
are feasible, by solving the parameterized problem of $\max_{\mathcal{P},\mathcal{S}}\mbox{ }\left\{ F(q_{i-1})\right\} $
at each iteration. Hence, it can be shown that the method produces
an increasing sequence of $q$ values, which converges to the optimal
value at a superlinear convergence rate. As shown in Table~\ref{algor:dinkelbach},
each outer iteration corresponds to solving $\max_{\mathcal{P},\mathcal{S}}\mbox{ }\left\{ F(q_{i-1})\right\} $,
where $q_{i-1}$ is a given value of the parameter $q$, to obtain
$\mathcal{P}_{i}^{*}$ and $\mathcal{S}_{i}^{*}$, which at the optimal
power and subcarrier values at the $i$th iteration of Dinkelbach's
method. For further details and a proof of convergence, please refer
to~\cite{Dinkelbach1967}.

\subsection{Solving the inner loop maximization problem\label{sub:Dinkelbach_inner}}

Dinkelbach's method relies on solving $\max_{\mathcal{P},\mathcal{S}}\mbox{ }F(q_{i-1})$,
in each iteration, which will henceforth be referred to as~(P$_{q_{i-1}}$).
Since it has been shown that $R_{T}(\mathcal{P},\mathcal{S})$ is
concave whilst $P_{T}(\mathcal{P},\mathcal{S})$ is affine, then~(P$_{q_{i-1}}$)
is concave w.r.t. the variables $\mathcal{P}$ and $\mathcal{S}$.
Assuming the existence of an interior point~(Slater's condition),
there is a zero duality gap between the dual problem of~(P$_{q_{i-1}}$)
and the primal problem of~(P)~\cite{Boyd2004}. Thus solving the
dual problem of~(P$_{q_{i-1}}$) is equivalent to solving the primal
problem of~(P).

\begin{figure*}[tbp]

\begin{eqnarray}
\mathcal{L}\left(\mathcal{P},\mathcal{S},\lambda\right) & = & \sum_{k=1}^{K}\sum_{n=1}^{N}s_{k}^{D,n}\mbox{log}_{2}\left(1+\frac{\widetilde{P}_{0,k}^{D,n}G_{0,k}^{n}}{s_{k}^{D,n}\Delta\gamma N_{0}W}\right)\nonumber \\
 &  & +\frac{s_{k}^{A,n}}{2}\mbox{log}_{2}\left(1+\frac{\widetilde{P}_{0,\mathcal{M}(k)}^{A,n}G_{0,\mathcal{M}(k)}^{n}\widetilde{P}_{\mathcal{M}(k),k}^{A,n}G_{\mathcal{M}(k),k}^{n}}{s_{k}^{A,n}\Delta\gamma N_{0}W\left(\widetilde{P}_{0,\mathcal{M}(k)}^{A,n}G_{0,\mathcal{M}(k)}^{n}+\widetilde{P}_{\mathcal{M}(k),k}^{A,n}G_{\mathcal{M}(k),k}^{n}\right)}\right)\nonumber \\
 &  & -q_{i-1}\left[\left(P_{C}^{(B)}+M\cdot P_{C}^{(R)}\right)+\sum_{k=1}^{K}\sum_{n=1}^{N}\xi^{(B)}\widetilde{P}_{0,k}^{D,n}+\frac{1}{2}\left(\xi^{(B)}\widetilde{P}_{0,\mathcal{M}(k)}^{A,n}+\xi^{(R)}\widetilde{P}_{\mathcal{M}(k),k}^{A,n}\right)\right]\nonumber \\
 &  & +\lambda\left(P_{max}-\sum_{k=1}^{K}\sum_{n=1}^{N}\widetilde{P}_{0,k}^{D,n}+\widetilde{P}_{0,\mathcal{M}(k)}^{A,n}+\widetilde{P}_{\mathcal{M}(k),k}^{A,n}\right).\label{eq:lagrange}
\end{eqnarray}
\hrulefill
\end{figure*}

The Lagrangian of~(P$_{q_{i-1}}$) is given by~(\ref{eq:lagrange}),
where $\lambda\geq0$ is the Lagrangian multiplier associated with
the constraint~(\ref{eq:C1}). The feasible region constraints~(\ref{eq:C4})
and~(\ref{eq:C5}), and constraints~(\ref{eq:C2}) and~(\ref{eq:C3})
will be considered when deriving the optimal solution, which is detailed
later.

The dual problem of~(P$_{q_{i-1}}$) may be written as~\cite{Palomar2006}
\begin{equation}
\underset{\lambda\geq0}{\mbox{min.}}\mbox{ }g(\lambda)=\underset{\lambda\geq0}{\mbox{min.}}\mbox{ }\underset{\mathcal{P},\mathcal{S}}{\mbox{max.}}\mbox{ }\mathcal{L}\left(\mathcal{P},\mathcal{S},\lambda\right),\label{eq:dual}
\end{equation}
which is solved using the dual decomposition approach~\cite{Palomar2006}.
Using dual decomposition,~(\ref{eq:dual}) may be readily solved
via solving $NK$ similar subproblems to obtain both the power as
well as subcarrier allocations, and by solving a master problem to
update $\lambda$. The dual decomposition approach is outlined in
the following.

\subsubsection{Solving the subproblem of power and subcarrier allocation}

For a fixed $\lambda$ and $q_{i-1}$, $\underset{\mathcal{P},\mathcal{S}}{\mbox{max.}}\mbox{ }\mathcal{L}\left(\mathcal{P},\mathcal{S},\lambda\right)$
is solved to obtain the corresponding power and subcarrier allocations.
Since the problem is now in a standard concave form, the Karush\textendash{}Kuhn\textendash{}Tucker~(KKT)
conditions~\cite{Boyd2004}, which are first-order necessary and
sufficient conditions for optimality, may be used in order to find
the optimal solution. All optimal variables are denoted by a superscript
asterisk. The total transmit power assigned for AF transmission to
user $k$ over subcarrier $n$ is now denoted by $\widetilde{P}_{k}^{A,n}=\widetilde{P}_{0,\mathcal{M}(k)}^{A,n}+\widetilde{P}_{\mathcal{M}(k),k}^{A,n}$.
Then, by substituting $\widetilde{P}_{\mathcal{M}(k),k}^{A,n}=\widetilde{P}_{k}^{A,n}-\widetilde{P}_{0,\mathcal{M}(k)}^{A,n}$
into~(\ref{eq:lagrange}), the following first-order derivatives
may be obtained
\begin{equation}
\left.\frac{\partial\mathcal{L}\left(\mathcal{P},\mathcal{S},\lambda\right)}{\partial\widetilde{P}_{0,k}^{D,n}}\right|_{\widetilde{P}_{0,k}^{D,n}=\widetilde{P}_{0,k}^{D,n*}}=0,\label{eq:dPb_D}
\end{equation}
\begin{equation}
\left.\frac{\partial\mathcal{L}\left(\mathcal{P},\mathcal{S},\lambda\right)}{\partial\widetilde{P}_{k}^{A,n}}\right|_{\widetilde{P}_{k}^{A,n}=\widetilde{P}_{k}^{A,n*}}=0\label{eq:dPt_A}
\end{equation}
and
\begin{equation}
\left.\frac{\partial\mathcal{L}\left(\mathcal{P},\mathcal{S},\lambda\right)}{\partial\widetilde{P}_{0,\mathcal{M}(k)}^{A,n}}\right|_{\widetilde{P}_{0,\mathcal{M}(k)}^{A,n}=\widetilde{P}_{0,\mathcal{M}(k)}^{A,n*}}=0.\label{eq:dPb_A}
\end{equation}
The optimal values of $\widetilde{P}_{0,k}^{D,n}$ may be readily
obtained from~(\ref{eq:dPb_D}) as
\begin{equation}
P_{0,k}^{D,n*}=\left[\frac{1}{\ln2\left(q_{i-1}\xi^{(B)}+\lambda\right)}-\frac{1}{\alpha_{k}^{D,n}}\right]^{+},\label{eq:dinkel_Pd}
\end{equation}
where the effective channel gain of the direct transmission is given
by
\begin{equation}
\alpha_{k}^{D,n}=\frac{G_{0,k}^{n}}{\Delta\gamma N_{0}W}
\end{equation}
and $[\cdot]^{+}$ denotes $\max(0,\cdot)$ since the powers allocated
have to be nonnegative due to the constraint~(\ref{eq:C4}). Similarly
the optimal values of $\widetilde{P}_{0,\mathcal{M}(k)}^{A,n}$ and
$\widetilde{P}_{\mathcal{M}(k),k}^{A,n}$ may be obtained by equating~(\ref{eq:dPt_A})
and~(\ref{eq:dPb_A}) to give
\begin{equation}
P_{0,\mathcal{M}(k)}^{A,n*}=\beta_{k}^{A,n}P_{k}^{A,n*}\label{eq:dinkel_Pb}
\end{equation}
and
\begin{equation}
P_{\mathcal{M}(k),k}^{A,n*}=\left(1-\beta_{k}^{A,n}\right)P_{k}^{A,n*},\label{eq:dinkel_Pr}
\end{equation}
where the total transmit power assigned for the AF transmission to
user $k$ over subcarrier $n$ is given by~(\ref{eq:dinkel_Pt_A}),~(\ref{eq:alpha_A})
and~(\ref{eq:beta_A}).\begin{figure*}[tbp]
\begin{equation}
P_{k}^{A,n*}=\left[\frac{1}{\ln2\left(\beta_{k}^{A,n}\left(q_{i-1}\xi^{(B)}+2\lambda\right)+\left(1-\beta_{k}^{A,n}\right)\left(q_{i-1}\xi^{(R)}+2\lambda\right)\right)}-\frac{1}{\alpha_{k}^{A,n}}\right]^{+}\label{eq:dinkel_Pt_A}
\end{equation}
\begin{equation}
\alpha_{k}^{A,n}=\frac{\beta_{k}^{A,n}\left(1-\beta_{k}^{A,n}\right)G_{0,\mathcal{M}(k)}^{n}G_{\mathcal{M}(k),k}^{n}}{\left(\beta_{k}^{A,n}G_{0,\mathcal{M}(k)}^{n}+\left(1-\beta_{k}^{A,n}\right)G_{\mathcal{M}(k),k}^{n}\right)\Delta\gamma N_{0}W}\label{eq:alpha_A}
\end{equation}
\begin{equation}
\beta_{k}^{A,n}=\frac{-G_{\mathcal{M}(k),k}^{n}\left(q_{i-1}\xi^{(R)}+2\lambda\right)+\sqrt{G_{0,\mathcal{M}(k)}^{n}G_{\mathcal{M}(k),k}^{n}\left(q_{i-1}\xi^{(B)}+2\lambda\right)\left(q_{i-1}\xi^{(R)}+2\lambda\right)}}{G_{0,\mathcal{M}(k)}^{n}\left(q_{i-1}\xi^{(B)}+2\lambda\right)-G_{\mathcal{M}(k),k}^{n}\left(q_{i-1}\xi^{(R)}+2\lambda\right)}\label{eq:beta_A}
\end{equation}
\hrulefill
\end{figure*} Observe that~(\ref{eq:beta_A}) is the fraction of the total AF
transmit power that is allocated for the BS-to-RN link while obeying
$0\leq\beta_{k}^{A,n}\leq1$.

Having calculated the optimal power allocations, the optimal subcarrier
allocations may be derived using the first-order derivatives as follows:
\begin{eqnarray}
\frac{\partial\mathcal{L}\left(\mathcal{P},\mathcal{S},\lambda\right)}{\partial s_{k}^{D,n}} & = & \log_{2}\left(1+\alpha_{k}^{D,n}P_{0,k}^{D,n*}\right)\nonumber \\
 &  & -\frac{\alpha_{k}^{D,n}P_{0,k}^{D,n*}}{\ln2\left(1+\alpha_{k}^{D,n}P_{0,k}^{D,n*}\right)}\nonumber \\
 & = & D_{k}^{n}\begin{cases}
<0 & \mbox{if }s_{k}^{D,n*}=0,\\
=0 & \mbox{if }s_{k}^{D,n*}\in(0,1)\\
>0 & \mbox{if }s_{k}^{D,n*}=1
\end{cases},\label{eq:dinkel_Dnk}
\end{eqnarray}
and
\begin{eqnarray}
\frac{\partial\mathcal{L}\left(\mathcal{P},\mathcal{S},\lambda\right)}{\partial s_{k}^{A,n}} & = & \frac{1}{2}\log_{2}\left(1+\alpha_{k}^{A,n}\widetilde{P}_{k}^{A,n*}\right)\nonumber \\
 &  & -\frac{\alpha_{k}^{A,n}\widetilde{P}_{k}^{A,n*}}{2\ln2\left(1+\alpha_{k}^{A,n}\widetilde{P}_{k}^{A,n*}\right)}\\
 & = & A_{k}^{n}\begin{cases}
<0 & \mbox{if }s_{k}^{A,n*}=0,\\
=0 & \mbox{if }s_{k}^{A,n*}\in(0,1)\\
>0 & \mbox{if }s_{k}^{A,n*}=1.
\end{cases},\label{eq:dinkel_Ank}
\end{eqnarray}
(\ref{eq:dinkel_Dnk}) and~(\ref{eq:dinkel_Ank}) stem from the fact
that if the optimal value of $s_{k}^{X,n}$ occurs at the boundary
of the feasible region, then $\mathcal{L}\left(\mathcal{P},\mathcal{S},\lambda\right)$
must be decreasing with the values of $s_{k}^{X,n}$ that approach
the interior of the feasible region. By contrast, for example, the
derivative $D_{k}^{n}=0$ if the optimal $s_{k}^{D,n}$ is obtained
in the interior of the feasible region~\cite{Wong1999}. However,
since each subcarrier may only be used for transmission to a single
user, each subcarrier $n$ is allocated to the specific user $k$
having the highest value of $\max\left(A_{k}^{n},D_{k}^{n}\right)$
in order to achieve the highest increase in $\mathcal{L}\left(\mathcal{P},\mathcal{S},\lambda\right)$.
The optimal allocation for subcarrier $n$ is as follows%
\footnote{If there are multiple users that tie for the value of $\max\left(A_{k}^{n},D_{k}^{n}\right)$,
then a random user from the maximal set is chosen.%
}
\begin{equation}
s_{k}^{D,n*}=\left\{ \begin{array}{cc}
1, & \mbox{if }D_{k}^{n}=\max_{j}\left[\max\left(A_{j}^{n},D_{j}^{n}\right)\right]\mbox{ and }D_{k}^{n}\geq0,\\
0, & \mbox{otherwise,}
\end{array}\right.\label{eq:dinkel_sd}
\end{equation}
and

\begin{equation}
s_{k}^{A,n*}=\left\{ \begin{array}{cc}
1, & \mbox{if }A_{k}^{n}=\max_{j}\left[\max\left(A_{j}^{n},D_{j}^{n}\right)\right]\mbox{ and }A_{k}^{n}\geq0,\\
0, & \mbox{otherwise.}
\end{array}\right.\label{eq:dinkel_sa}
\end{equation}
Thus constraints~(\ref{eq:C2})-~(\ref{eq:C5}) are satisfied and
the optimal primal variables are obtained for a given $\lambda$.
Observe that the optimal power allocations given by~(\ref{eq:dinkel_Pd})
and~(\ref{eq:dinkel_Pt_A}) are indeed customized water-filling solutions~\cite{Goldsmith2005},
where the effective channel gains are given by $\alpha_{k}^{D,n}$
and $\alpha_{k}^{A,n}$, respectively, and where the water levels
are determined both by the cost of allocating power, $\lambda$, as
well as the current cost of power to the EE given by $q_{i-1}$.

\subsubsection{Updating the dual variable $\lambda$}

Since~(\ref{eq:dinkel_Pd}),~(\ref{eq:dinkel_Pb}),~(\ref{eq:dinkel_Pr}),~(\ref{eq:dinkel_sd})
and~(\ref{eq:dinkel_sa}) give a unique solution for $\mbox{\ensuremath{\underset{\mathcal{P},\mathcal{S}}{\mbox{max.}}} }\mathcal{L}\left(\mathcal{P},\mathcal{S},\lambda\right)$,
it follows that $g(\lambda)$ is differentiable and hence the gradient
method~\cite{Boyd2004,Palomar2006} may be readily used for updating
the dual variables $\lambda$. The gradient of $\lambda$ is given
by
\begin{eqnarray}
\frac{\partial\mathcal{L}\left(\mathcal{P},\mathcal{S},\lambda\right)}{\partial\lambda} & = & P_{max}-\sum_{k=1}^{K}\sum_{n=1}^{N}\Big(\widetilde{P}_{0,k}^{D,n}\nonumber \\
 &  & +\widetilde{P}_{0,\mathcal{M}(k)}^{A,n}+\widetilde{P}_{\mathcal{M}(k),k}^{A,n}\Big).
\end{eqnarray}
Therefore, $\lambda$ may be updated using the optimal variables to
give~(\ref{eq:Dinkel_lambda}),\begin{figure*}[tbp]
\begin{equation}
\lambda(i+1)=\left[\lambda(i)-\alpha_{\lambda}(i)\left(P_{max}-\sum_{k=1}^{K}\sum_{n=1}^{N}\widetilde{P}_{0,k}^{D,n*}+\widetilde{P}_{0,\mathcal{M}(k)}^{A,n*}+\widetilde{P}_{\mathcal{M}(k),k}^{A,n*}\right)\right]^{+}\label{eq:Dinkel_lambda}
\end{equation}
\hrulefill
\end{figure*} where $\alpha_{\lambda}(i)$ is the size of the step taken in the
direction of the negative gradient for the dual variable $\lambda$
at iteration $i$. For the performance investigations of Section~\ref{sec:Results-and-Discussions},
a constant step size is used, since it is comparatively easier to
find a value that strikes a balance between optimality and convergence
speed. The process of computing the optimal power as well as subcarrier
allocations and subsequently updating $\lambda$ is repeated until
convergence is attained, indicating that the dual optimal point has
been reached. Since the primal problem~(P$_{q_{i-1}}$) is concave,
there is zero duality gap between the dual and primal solutions. Hence,
solving the dual problem is equivalent to solving the primal problem.
The inner loop solution method is summarized in Table~\ref{algor:dinkelbach_inner}.
\begin{table}
\setlength{\arrayrulewidth}{1pt}

\centering{}\caption{Inner loop solution method for obtaining the optimal power and subcarrier
allocations for a given $q_{i-1}$.}
\label{algor:dinkelbach_inner}%
\begin{tabular}{rl}
\hline 
\multicolumn{2}{l}{\textbf{Algorithm 2} Inner loop solution method for obtaining the
optimal power}\tabularnewline
\multicolumn{2}{l}{and subcarrier allocations for a given $q_{i-1}$.}\tabularnewline
\hline 
Input: & $I_{inner}^{D}$~(maximum number of iterations)\tabularnewline
 & $\epsilon_{inner}^{D}>0$~(convergence tolerance)\tabularnewline
 & end do\tabularnewline
1: & $i\leftarrow0$\tabularnewline
2: & \textbf{do while} $|\lambda(i)-\lambda(i-1)|>\epsilon_{inner}^{D}$
\textbf{and} $i<I_{inner}^{D}$\tabularnewline
3: & \quad{}$i\leftarrow i+1$\tabularnewline
4: & \quad{}\textbf{for} $n$ \textbf{from} $1$ \textbf{to} $N$\tabularnewline
5: & \quad{}\quad{}\textbf{for} $k$ \textbf{from} $1$ \textbf{to}
$K$\tabularnewline
6: & \quad{}\quad{}\quad{}Obtain the optimal power allocation using~(\ref{eq:dinkel_Pd}),~(\ref{eq:dinkel_Pb})\tabularnewline
 & \quad{}\quad{}\quad{}and~(\ref{eq:dinkel_Pr})\tabularnewline
7: & \quad{}\quad{}\textbf{end for}\tabularnewline
8: & \quad{}\quad{}Obtain the optimal subcarrier allocation using~(\ref{eq:dinkel_sd})
and~(\ref{eq:dinkel_sa})\tabularnewline
9: & \quad{}\textbf{end for}\tabularnewline
10: & \quad{}Update the dual variables $\lambda$ using~(\ref{eq:Dinkel_lambda})\tabularnewline
11: & \textbf{end do}\tabularnewline
12: & \textbf{return}\tabularnewline
\hline 
\end{tabular}
\end{table}

\subsection{Summary of solution methodology}

\begin{figure}[h]
\begin{centering}
\includegraphics[scale=0.7]{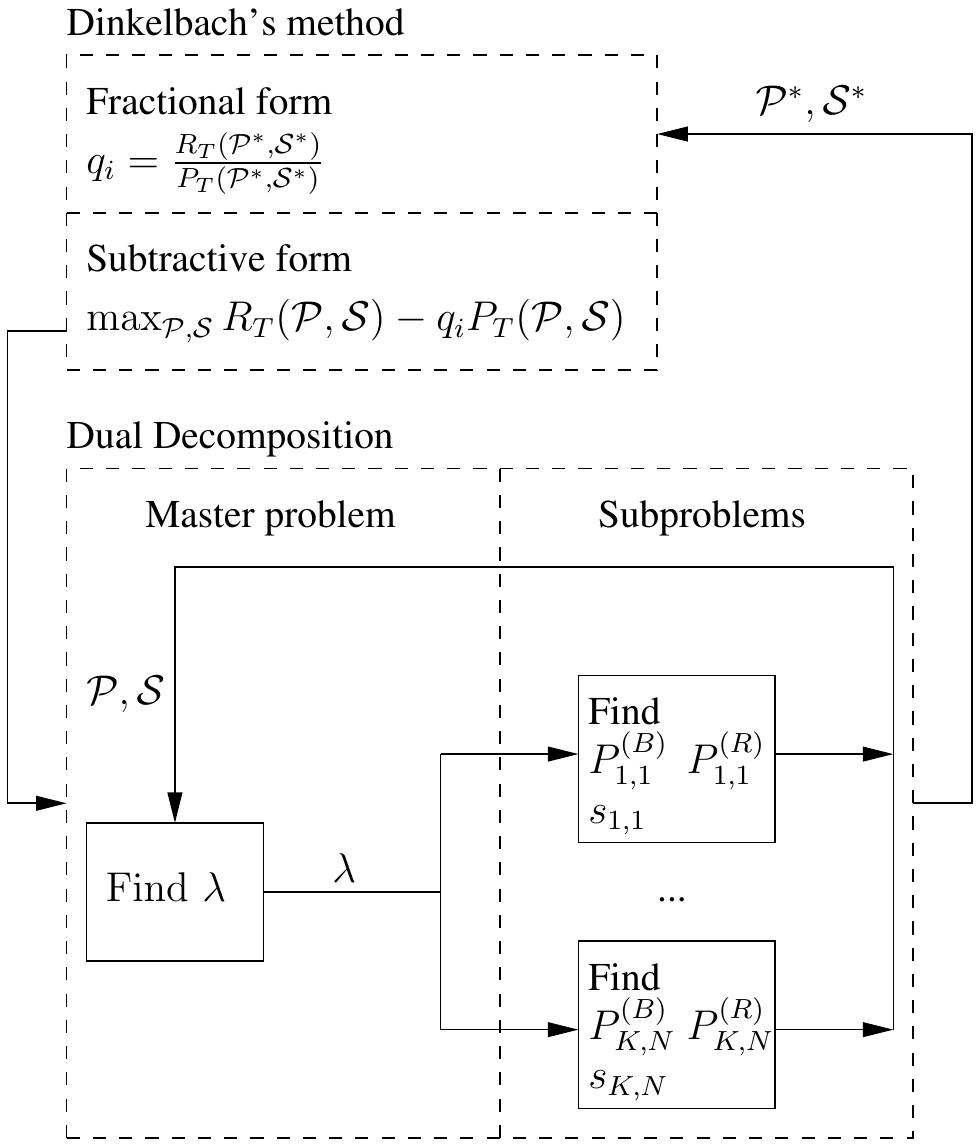}
\par\end{centering}

\caption{Summary of the solution methodology for the relaxed problem (P).}
\label{fig:summary}
\end{figure}
Again, for additional clarity, the solution methodology is summarized
in Fig.~\ref{fig:summary}. Firstly, the relaxed problem~(P) expressed
in a fractional form is rewritten as a subtractive, parameterized
concave form, where $q_{i}$ is the parameter. Solving this problem
for a given $q_{i}$ is termed an outer iteration, which is described
in Section~\ref{sub:Dinkelbach_outer} and is illustrated in the
upper block of Fig.~\ref{fig:summary}. The lower block of Fig.~\ref{fig:summary}
illustrates the dual decomposition approach that is employed for solving
the subtractive, concave problem. Each iteration of the dual decomposition
approach is termed an inner iteration, which is further detailed in
Section~\ref{sub:Dinkelbach_inner}. In each inner iteration, $2NK$
subproblems are solved to obtain the optimal power and subcarrier
variables for a given $q_{i}$ and for the dual variable $\lambda$.
The dual variable $\lambda$ is then updated, depending on the power
and subcarrier variables obtained. Multiple inner iterations are completed
until convergence of the optimal dual and primal solutions is reached.
The optimal $\mathcal{P}^{*}$ and $\mathcal{S}^{*}$ are then fed
back into the upper block~\ref{fig:summary} to evaluate the updated
value of $q_{i}$, which is used in the next outer iteration. Several
outer iterations are completed until convergence to the optimal $q_{i}$
is obtained. The corresponding optimal $\mathcal{P}^{*}$ and $\mathcal{S}^{*}$
values are the power and subcarrier allocation variables that solve
the problem~(P). The algorithmic complexity of this method is dominated
by the comparison operations given by~(\ref{eq:dinkel_sd}) and~(\ref{eq:dinkel_sa}),
which leads to a total complexity of $\mathcal{O}\left(I_{dual}\times2NK\right)$
when $NK$ is large, where $I_{dual}$ is the total number of inner
iterations required for reaching convergence in Dinkelbach's method.

\section{Results and Discussions\label{sec:Results-and-Discussions}}

\begin{table}
\centering{}\caption{Simulation parameters used to obtain all results in this section unless
otherwise specified.}
\label{tab:param}%
\begin{tabular}{|l|r|}
\hline 
Simulation parameter & Value\tabularnewline
\hline 
\hline 
Subcarrier bandwidth, $W$ Hertz & $12$k\tabularnewline
\hline 
Number of RNs, $M$ & $\{0,1,2,3,5,6\}$\tabularnewline
\hline 
Number of subcarriers, $N$ & $\{128,512,1024\}$\tabularnewline
\hline 
Number of UEs, $K$ & $\{30,60,120\}$\tabularnewline
\hline 
Cell radius, km & $\{0.75,1,1.25,1.5,1.75,2\}$\tabularnewline
\hline 
Ratio of BS-to-RN distance to the cell & $\{0.1,0.3,0.5,0.7,0.9\}$\tabularnewline
radius, $D_{r}$ & \tabularnewline
\hline 
SNR gap of wireless transceivers, & 0\tabularnewline
$\Delta\gamma$ dB & \tabularnewline
\hline 
\multicolumn{1}{|l|}{Maximum total transmission power,} & $\{0,5,10,15,20,25,30,$\tabularnewline
$P_{max}$ dBm & $35,40,45,50,55,60\}$\tabularnewline
\hline 
Fixed power consumption of the BS, & 60\tabularnewline
$P_{C}^{(B)}$ Watts~\cite{Arnold2010} & \tabularnewline
\hline 
Fixed power consumption of RNs, & 20\tabularnewline
$P_{C}^{(R)}$ Watts~\cite{Arnold2010} & \tabularnewline
\hline 
Reciprocal of the BS power amplifier's & 2.6\tabularnewline
 drain efficiency, $\xi^{(B)}$~\cite{Arnold2010} & \tabularnewline
\hline 
Reciprocal of the RNs' power amplifier's & 5\tabularnewline
 drain efficiency, $\xi^{(R)}$~\cite{Arnold2010} & \tabularnewline
\hline 
Noise power spectral density, & \textminus{}174\tabularnewline
$N_{0}$ dBm/Hz & \tabularnewline
\hline 
Maximum number of outer iterations in & 10\tabularnewline
Dinkelbach's algorithm, $I_{outer}^{D}$ & \tabularnewline
\hline 
Maximum number of inner iterations in & 100\tabularnewline
Dinkelbach's algorithm, $I_{inner}^{D}$ & \tabularnewline
\hline 
Convergence tolerance of iterative & $10^{-8}$\tabularnewline
 algorithms, $\epsilon_{outer}^{D}=\epsilon_{inner}^{D}$ & \tabularnewline
\hline 
Number of channel samples & $10^{4}$\tabularnewline
\hline 
\end{tabular}\vspace{-0.6cm}
\end{table}
This section presents the results of applying the EEM algorithm described
in Section~\ref{sec:Dinkelbach} to the relay-aided cellular system
shown in Fig.~\ref{fig:cellular}. Again, the channel is modeled
by the path-loss~\cite{3GPP_PL} and uncorrelated Rayleigh fading
obeying the complex normal distribution, $\mathcal{CN}(0,1)$. It
is assumed that the BS-to-RN link has line-of-sight~(LOS) propagation,
implying that a RN was placed on a tall building. However, the BS-to-UE
and RN-to-UE links typically have no LOS, since these links are likely
to be blocked by buildings and other large obstructing objects. The
RNs are evenly distributed at a fixed distance around the central
BS and the UEs are uniformly distributed within the cell. An independently-random
set of UE locations as well as fading channel realizations are generated
for each channel sample. For fair comparisons, the metrics used are
the average SE per subcarrier and the average EE per subcarrier. On
the other hand, the sum-rate may be calculated by multiplying the
average SE by $NW.$ Additionally, $\rho$ is introduced to denote
the average fraction of the total number of subcarriers that are used
for AF transmission. Thus, $\rho$ quantifies the benefit attained
from introducing RNs into the system. The main simulation parameters
are given in Table~\ref{tab:param}%
\footnote{For simplification, it is assumed that near-capacity transceivers
are employed in the network.%
}$^{,}$%
\footnote{In all cases, the step size and the initial value of $\lambda$ was
empirically optimized to give the optimal objective value in as few
iterations as possible, although the exact analytical method for achieving
this still remains an open issue.%
}.

\subsection{Convergence of iterative algorithms to optimal value}

\begin{figure}
\begin{centering}
\includegraphics[scale=0.8]{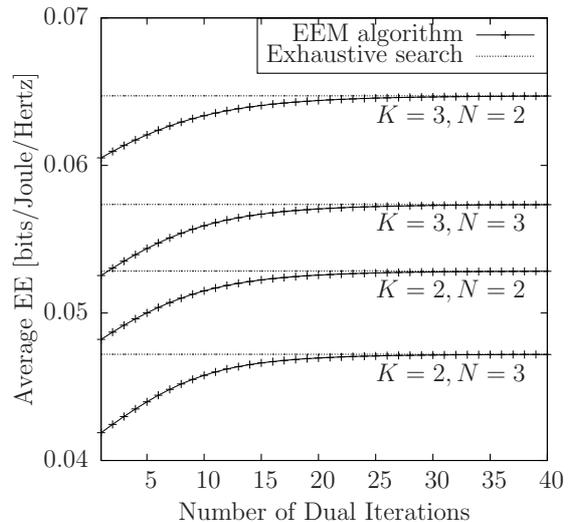}
\par\end{centering}

\caption{Average EE versus the total number of inner iterations of Dinkelbach's
method required for reaching convergence when using the simulation
parameters from Table~\ref{tab:param} with $P_{max}=0$dBm, $D_{r}=0.5$,
$M=0$ and with a cell radius of $1$km.}
\label{fig:conv_main}
\end{figure}
Fig.~\ref{fig:conv_main} illustrates the convergence behavior of
Dinkelbach's method invoked for maximizing the EE for a selection
of small-scale systems, averaged over $10^{4}$ different channel
realizations. Since the problem size is small, it is possible to generate
also the exhaustive-search based solution within a reasonable computation
time. As seen in Fig.~\ref{fig:conv_main}, Dinkelbach's method converges
to the optimal value within forty inner iterations. This result demonstrates
that the EEM algorithm based on Dinkelbach's method indeed obtains
the optimal power and subcarrier allocation, even though the relaxed
problem is solved and a high receiver's SNR was assumed.

\subsection{Effect of the number of UEs on the attainable SE and EE}

\begin{figure}
\begin{centering}
\subfloat[Average SE and $\rho$ versus $P_{max}$ for $K=30$, $60$ and $120$.]{\begin{centering}
\includegraphics[scale=0.8]{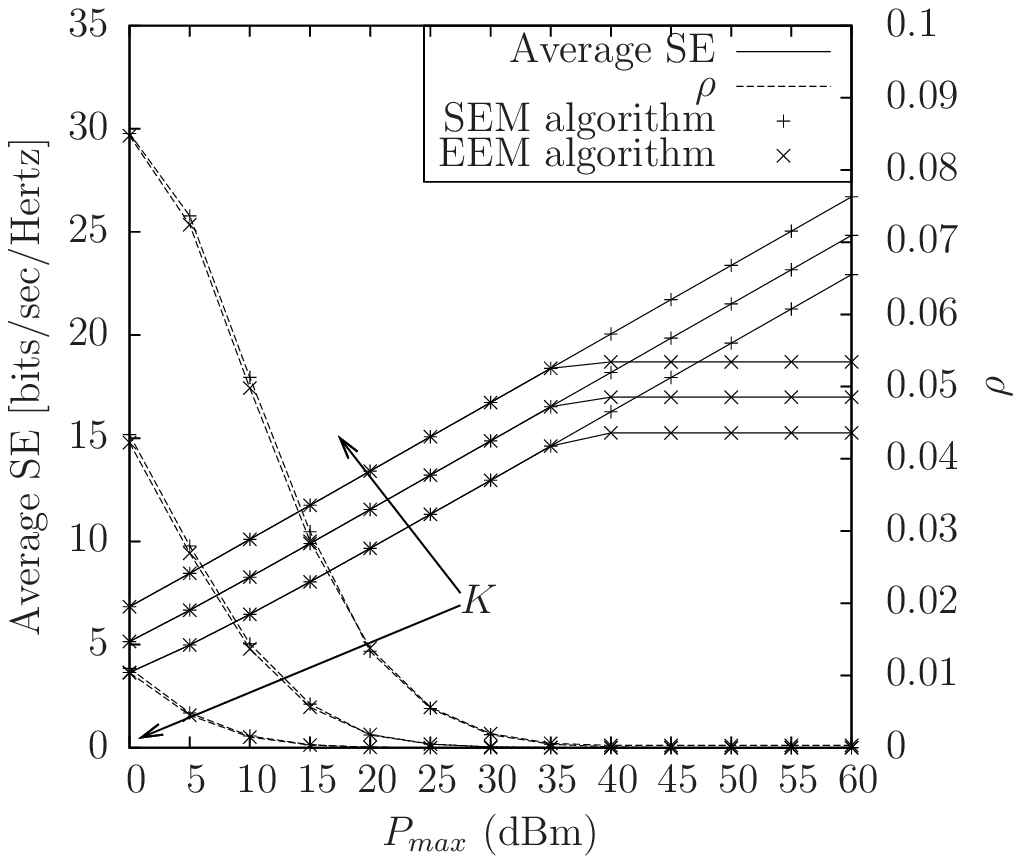}
\par\end{centering}

\label{fig:users_SE}}
\par\end{centering}

\begin{centering}
\subfloat[Average EE and $\rho$ versus $P_{max}$ for $K=30$, $60$ and $120$.]{\begin{centering}
\includegraphics[scale=0.8]{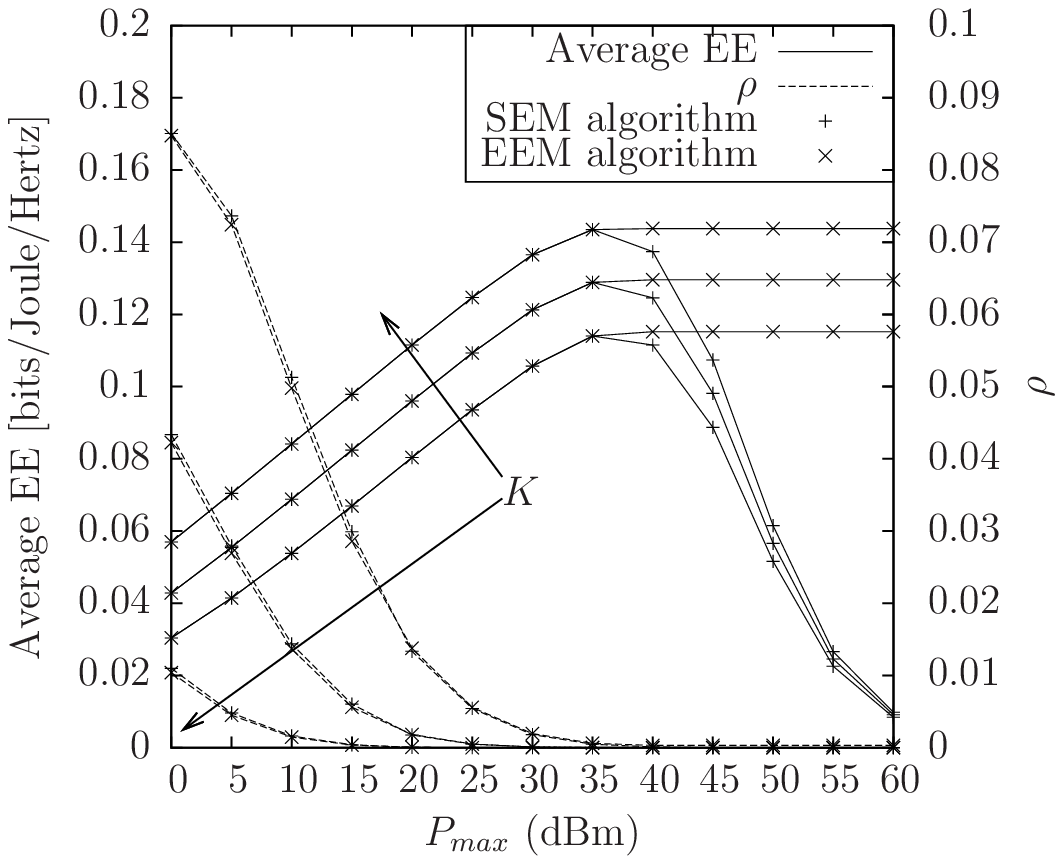}
\par\end{centering}

\label{fig:users_EE}}
\par\end{centering}

\centering{}\caption{Average SE, EE and $\rho$, and the effect of an increasing number
of users, $K$, for a system with simulation parameters from Table~\ref{tab:param}
with $N=128$, $M=3$, $D_{r}=0.5$ and with a cell radius of $1.5$km.}
\label{fig:users}
\end{figure}
Additionally, the EEM algorithm may be employed for evaluating the
effects of system-level design choices on the network's SE and EE.
The effect of $K$ on the average EE and SE%
\footnote{N.B. The maximum SE is obtained in the first outer iteration of Dinkelbach's
method with $q_{0}=0$, since this equates to zero penalty for any
power consumption.%
} is depicted in Fig.~\ref{fig:users}. As expected, upon increasing
$K$, the multi-user diversity of the system is increased, since the
scheduler is allowed to choose its subcarrier allocations from a larger
pool of channel gains. This results in an increase of both the maximum
EE as well as of the SE attained. Furthermore, Fig.~\ref{fig:users}
shows that as $P_{max}$ is increased, the SEM algorithm continues
to allocate more power in order to achieve a higher average SE at
the cost of EE, while the EEM algorithm attains the maximum EE and
does not continue to increase its attainable SE by sacrificing the
achieved EE. On the other hand, $\rho$ is inversely proportional
to $K$. This indicates that as the multi-user diversity increases,
the subcarriers are less likely to be allocated for AF transmissions,
simply because there are more favorable BS-to-UE channels owing to
having more UEs nearer to the cell-center. Moreover, the value of
$\rho$ decreases as $P_{max}$ increases, because there is more power
to allocate to the BS-to-UE links for UEs near the cell-center, which
benefit from a reduced pathloss as well as from a more efficient power
amplifier at the BS.

\subsection{Effect of the number of subcarriers on the attainable SE and EE}

\begin{figure}
\begin{centering}
\subfloat[Average SE and $\rho$ versus $P_{max}$ for $N=128$, $512$ and
$1024$.]{\begin{centering}
\includegraphics[scale=0.8]{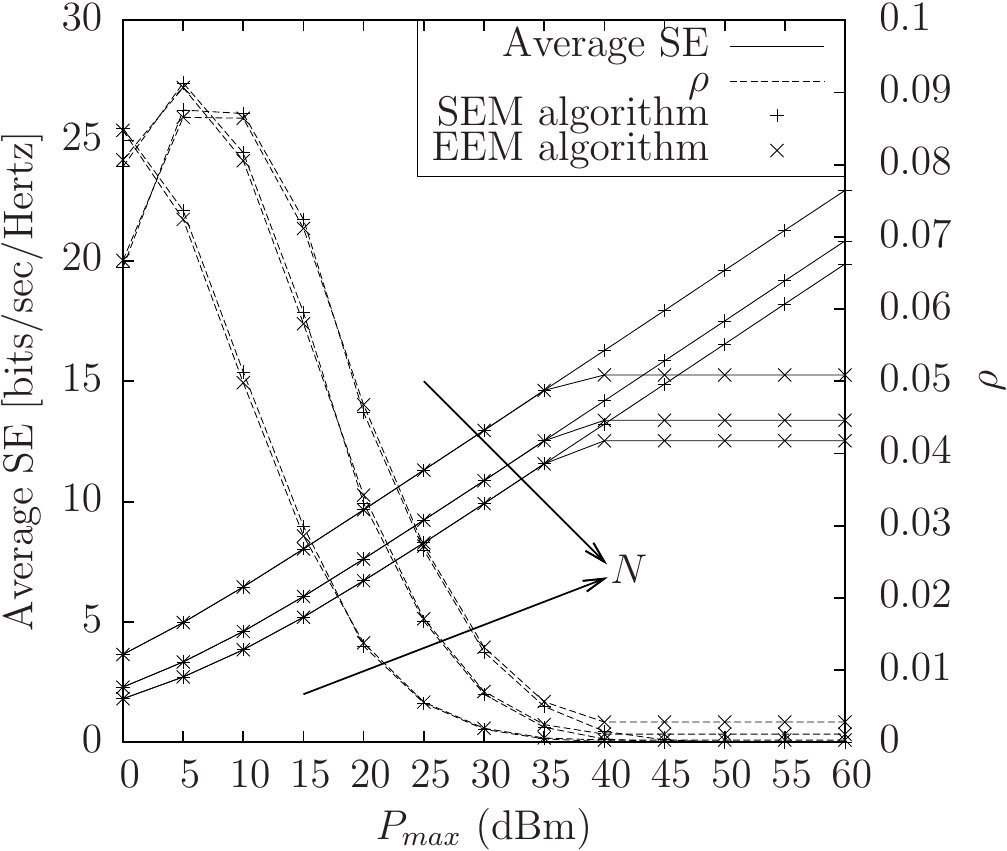}
\par\end{centering}

\label{fig:subcarriers_SE}}
\par\end{centering}

\centering{}\subfloat[Average EE and $\rho$ versus $P_{max}$ for $N=128$, $512$ and
$1024$.]{\begin{centering}
\includegraphics[scale=0.8]{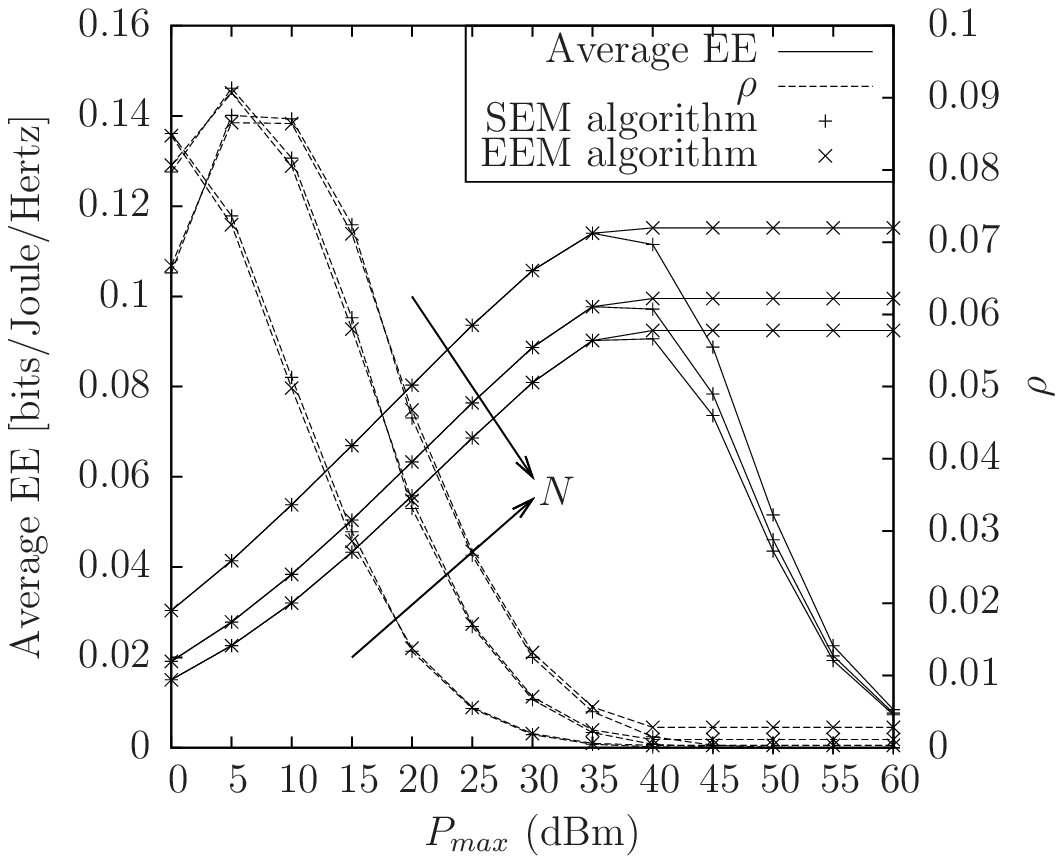}
\par\end{centering}

\label{fig:subcarriers_EE}}\caption{Average SE, EE and $\rho$, and the effect of an increasing number
of subcarriers, $N$, for a system with simulation parameters from
Table~\ref{tab:param} with $K=30$, $M=3$, $D_{r}=0.5$ and with
a cell radius of $1.5$km.}
\label{fig:subcarriers}
\end{figure}
Fig.~\ref{fig:subcarriers} illustrates the effect of increasing
$N$ on the attainable SE and EE. Fig.~\ref{fig:subcarriers} shows
much of the same trends as Fig.~\ref{fig:users}. For example, the
SE continues to rise at the cost of EE, as $P_{max}$ increases when
using the SEM algorithm, while the EEM algorithm attains the maximum
EE and the corresponding SE. However, in Fig.~\ref{fig:subcarriers}
both the SE and EE, averaged over $N$, decreases upon increasing
$N$, which implies that the subcarriers are not used effectively,
when more are available. On the other hand, it may be observed that
the sum-rate achieved using the SEM algorithm increases with $N$,
as expected due to frequency diversity.

In Fig.~\ref{fig:subcarriers}, it is noticeable that $\rho$ increases
upon increasing $N$, which is in contrast to the trend observed in
Fig.~\ref{fig:users}. This may be understood by considering the
UEs within the network. Since the UEs positions are fixed, as $N$
increases the scheduler has access to a larger pool of channel gains
for each individual UE, thus more cell-edge users may be supported
for maximizing the SE or EE. However, increasing $K$ does not have
this effect, since a larger $K$ value indicates that there are more
UEs near the cell-center, and since both the SEM and EEM algorithms
are greedy, these cell-center UEs are served before the cell-edge
UEs, hence $\rho$ decreases. The reduction of $\rho$ when $P_{max}$
is very small suggests that the total available power in the system
is not high enough to take advantage of the AF transmissions.

\subsection{Effect of the cell radius on the attainable SE and EE}

\begin{figure}
\begin{centering}
\subfloat[Average SE and $\rho$ versus $P_{max}$ for $M=0$, $1$, $2$, $3$,
$5$ and $6$.]{\begin{centering}
\includegraphics[scale=0.8]{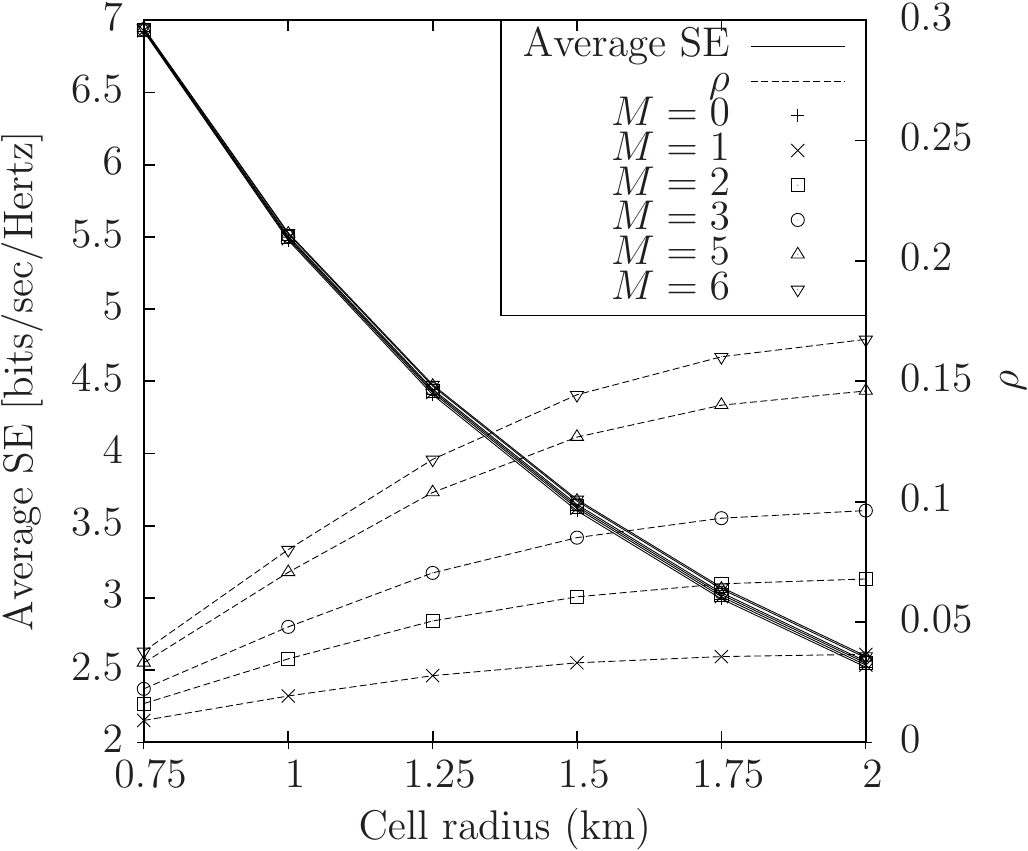}
\par\end{centering}

\label{fig:ISD_SE}}
\par\end{centering}

\centering{}\subfloat[Average EE and $\rho$ versus $P_{max}$ for $M=0$, $1$, $2$, $3$,
$5$ and $6$.]{\begin{centering}
\includegraphics[scale=0.8]{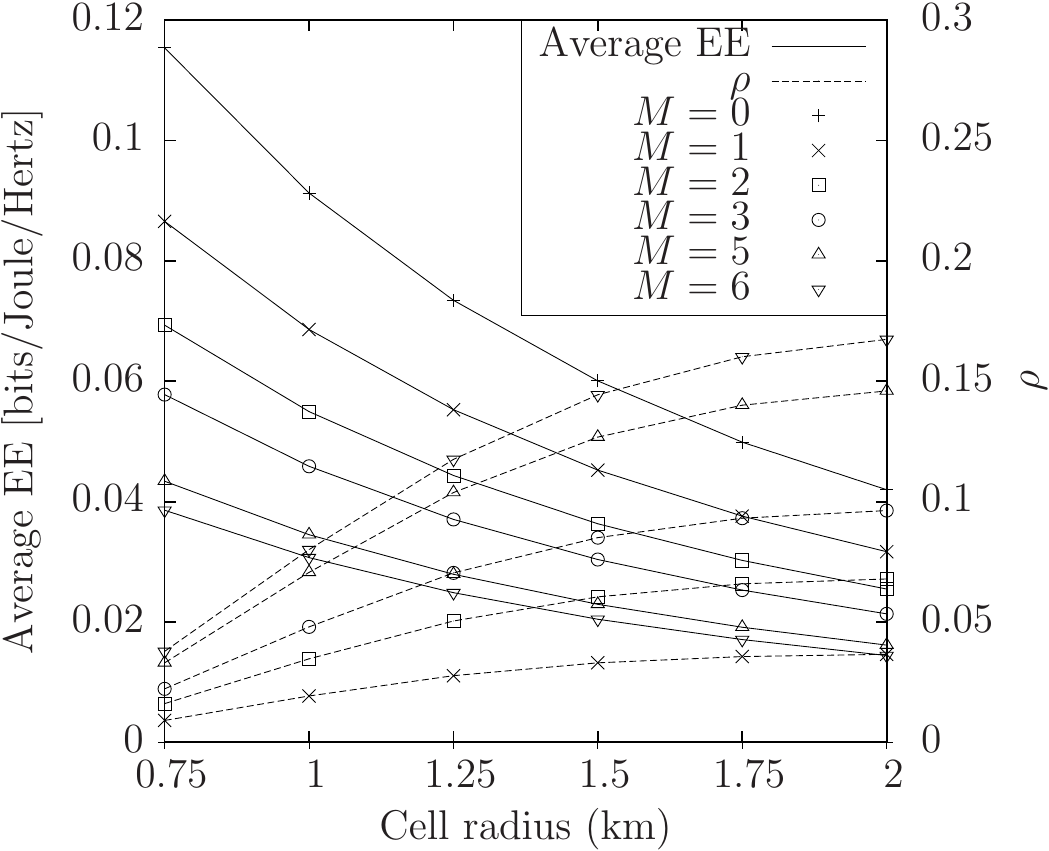}
\par\end{centering}

\label{fig:ISD_EE}}\caption{Average SE, EE and $\rho$, and the effect of an increasing cell radius
for a system with simulation parameters from Table~\ref{tab:param}
with $K=30$, $N=128$, $M\in\{0,1,2,3,5,6\}$, $D_{r}=0.5$ and with
a $P_{max}=0$dBm.}
\label{fig:ISD}
\end{figure}
The effect of increasing the cell radius on the attainable SE and
EE is characterized in Fig.~\ref{fig:ISD}. As expected, increasing
the cell radius has a detrimental effect on both the SE and EE of
the system regardless of the number of RNs employed owing to the increased
pathlosses experienced. Additionally, it is noteworthy that $\rho$
increases as the cell radius increases, indicating that relaying is
more beneficial for larger cells. On the other hand, the increase
in the SE obtained from employing RNs in a large cell is small. For
example, the SE is improved by a factor of $1.03$ when $M=6$ RNs
are used instead of $M=0$ at a cell radius of $2$km. This improvement
is modest when compared to the reduction in EE of a factor of $0.34$
due to having to support additional transmitting entities. This phenomenon
stems from the fact that, since the UEs are uniformly distributed
across the cell, the UEs nearer the cell-center are more likely to
be allocated resources for maximizing the SE or EE as they may benefit
from the more-favorable direct transmission. Thus, increasing the
number of RNs in the system brings a marginal benefit in terms of
SE or EE.

\subsection{Effect of the relay's position on the attainable SE and EE}

\begin{figure}
\begin{centering}
\subfloat[Average SE and $\rho$ versus $P_{max}$ for $M=0$, $1$, $2$, $3$,
$5$ and $6$.]{\begin{centering}
\includegraphics[scale=0.8]{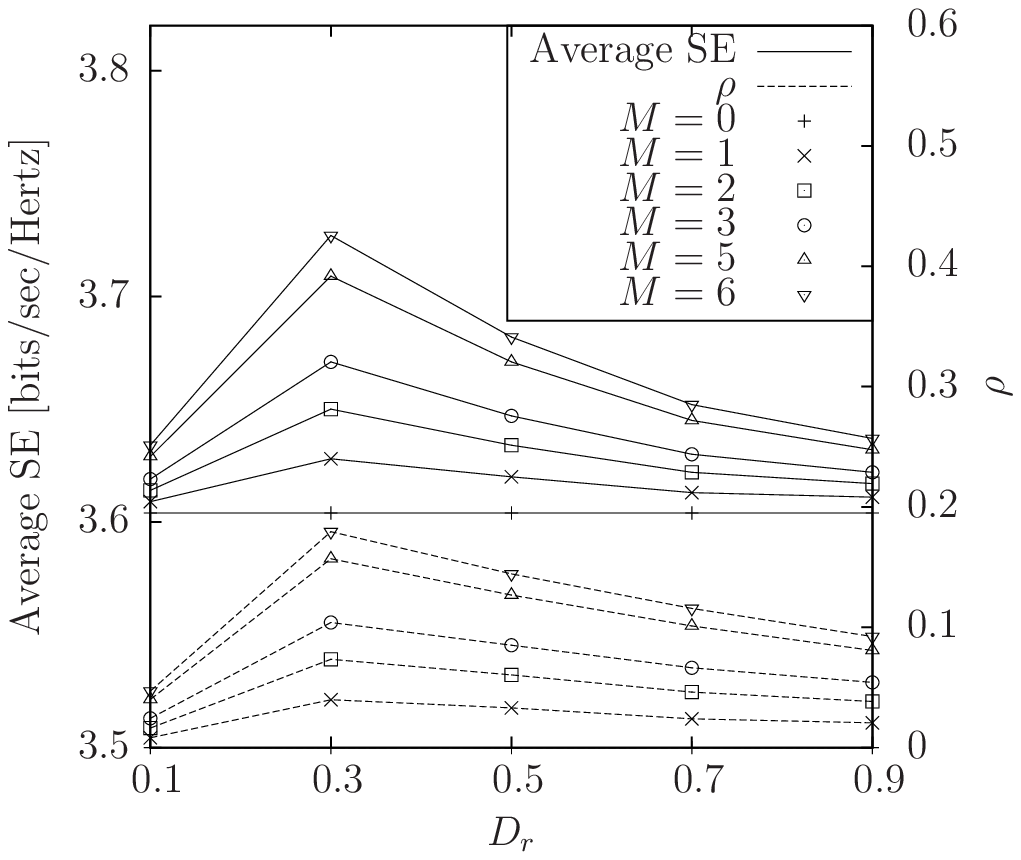}
\par\end{centering}

\label{fig:D_R_SE}}
\par\end{centering}

\centering{}\subfloat[Average EE and $\rho$ versus $P_{max}$ for $M=0$, $1$, $2$, $3$,
$5$ and $6$.]{\begin{centering}
\includegraphics[scale=0.8]{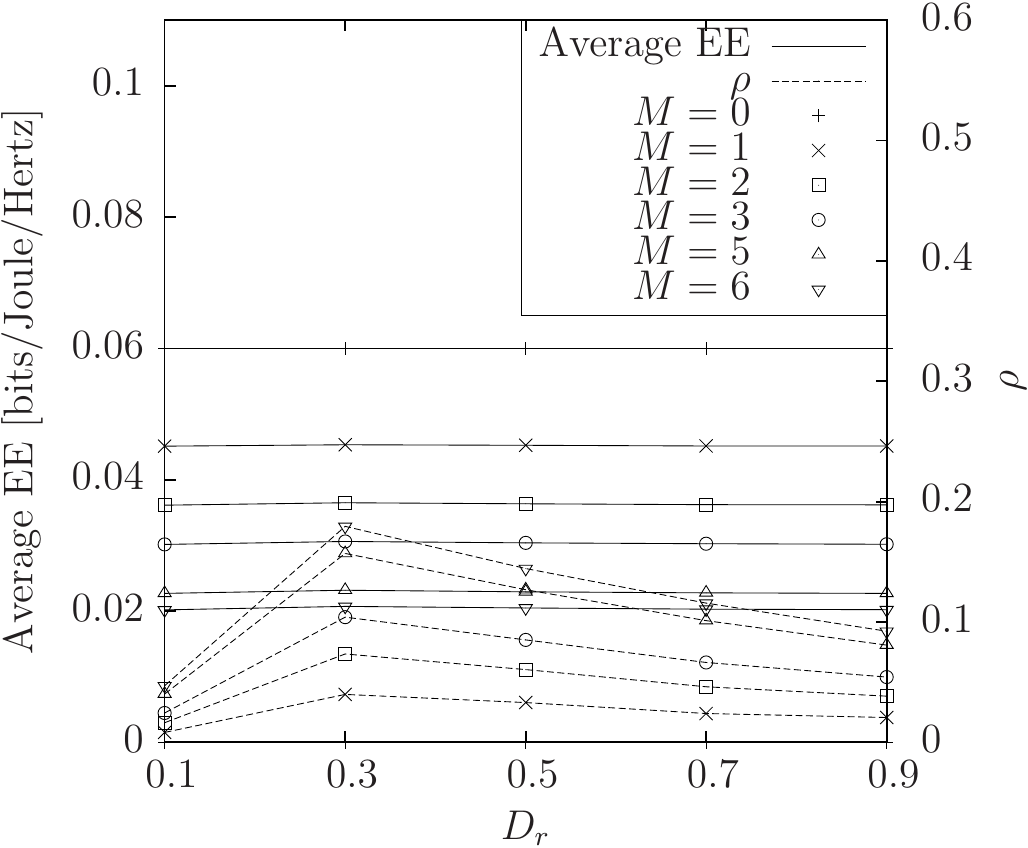}
\par\end{centering}

\label{fig:D_R_EE}}\caption{Average SE, EE and $\rho$, and the effect of an increasing $D_{r}$,
with simulation parameters from Table~\ref{tab:param} with $K=30$,
$N=128$, $M\in\{0,1,2,3,5,6\}$, $P_{max}=0$dBm and with a cell
radius of $1.5$km.}
\label{fig:D_R}
\end{figure}
The effect of the RNs' position relative to the BS and the cell-edge
is illustrated in Fig.~\ref{fig:D_R}, which clearly shows that the
optimal SE and EE is obtained, when the RN is closer to the BS than
to the UEs. This stems from the fact that the RN benefits from having
a stronger LOS link to the BS, when it is placed closer to the BS,
thus strengthening the AF links. However, it cannot be placed too
close to the BS, since the benefits gleaned from having a stronger
BS-to-RN link are then outweighed by having a more hostile RN-to-UE
link.

\section{Conclusions\label{sec:Conclusions}}

In this paper, the joint power and subcarrier allocation problem was
formulated for maximizing the EE in a multi-relay aided multi-user
OFDMA cellular network. The OF was proven to be quasi-concave and
an iterative method, namely Dinkelbach's method, was employed for
solving the associated optimization. Dinkelbach's method solves the
optimization problem by solving a sequence of subtractive concave
problems, which were solved using the dual decomposition approach
in this paper. The optimal power and subcarrier allocation were presented
for solving each iteration of the dual decomposition algorithm, and
simulations were performed to validate the algorithm.

Further simulation results show that when there is insufficient power
for attaining the maximum achievable EE, both the EEM and the SEM
algorithms have the same solution. As the system's total power is
increased, the SEM algorithm continues to allocate more power in order
to achieve ever higher values of SE, whereas the EEM algorithm reaches
an upper bound and does not make use of the additional available power.
Additionally, a comprehensive study of the effect of various system
parameters on the achievable SE and EE is performed. To summarize,
the achievable SE and EE is increased, when there is a larger number
of UEs in the system owing to achieving a higher multi-user diversity.
Increasing the number of available subcarriers, although increases
the SR owing to frequency diversity, reduces the average SE since
not all subcarriers are effectively utilized. The benefit of introducing
RNs into the network for improving the achievable SE becomes more
significant as the cell-size increases or the number of relays increases.
However, the EE is then degraded due to the increased overhead power
consumption. Furthermore, relaying is more beneficial, when the RNs
are placed closer to the BS, if there exists a LOS link between the
RNs and BS.

As a next step, EEM and SEM algorithms will be invoked for multi-cell
systems, which are interference-limited, rather than noise-limited.
This is in contrast to this work, which stipulated the idealized simplifying
assumption that the inter-cell interference is sufficiently low, and
thus may be ignored. Furthermore, on-line near-real-time optimization
for mobile RNs may be considered.\bibliographystyle{IEEEtran}
\bibliography{references}
\begin{IEEEbiography}
[{\includegraphics[width=1in,height=1.25in,clip,keepaspectratio]{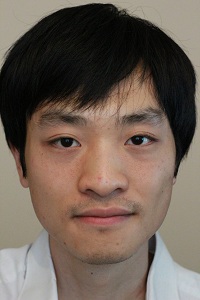}}]{Kent Tsz Kan Cheung} (S'09) received his B.Eng. degree (first-class honors) in electronic engineering from the Univeristy of Southampton, Southampton, U.K., in 2009. Since then he has been working towards a Ph.D. degree in wireless communications at the same institution. He was a recipient of the EPSRC Industrial CASE award in 2009, and was involved with the Core 5 Green Radio project of the Virtual Centre of Excellence in Mobile and Personal Communications (Mobile VCE).

His research interests include energy-efficiency, multi-carrier MIMO communications, cooperative communications, resource allocation and optimization.
\end{IEEEbiography}

\begin{IEEEbiography}
[{\includegraphics[width=1in,height=1.25in,clip,keepaspectratio]{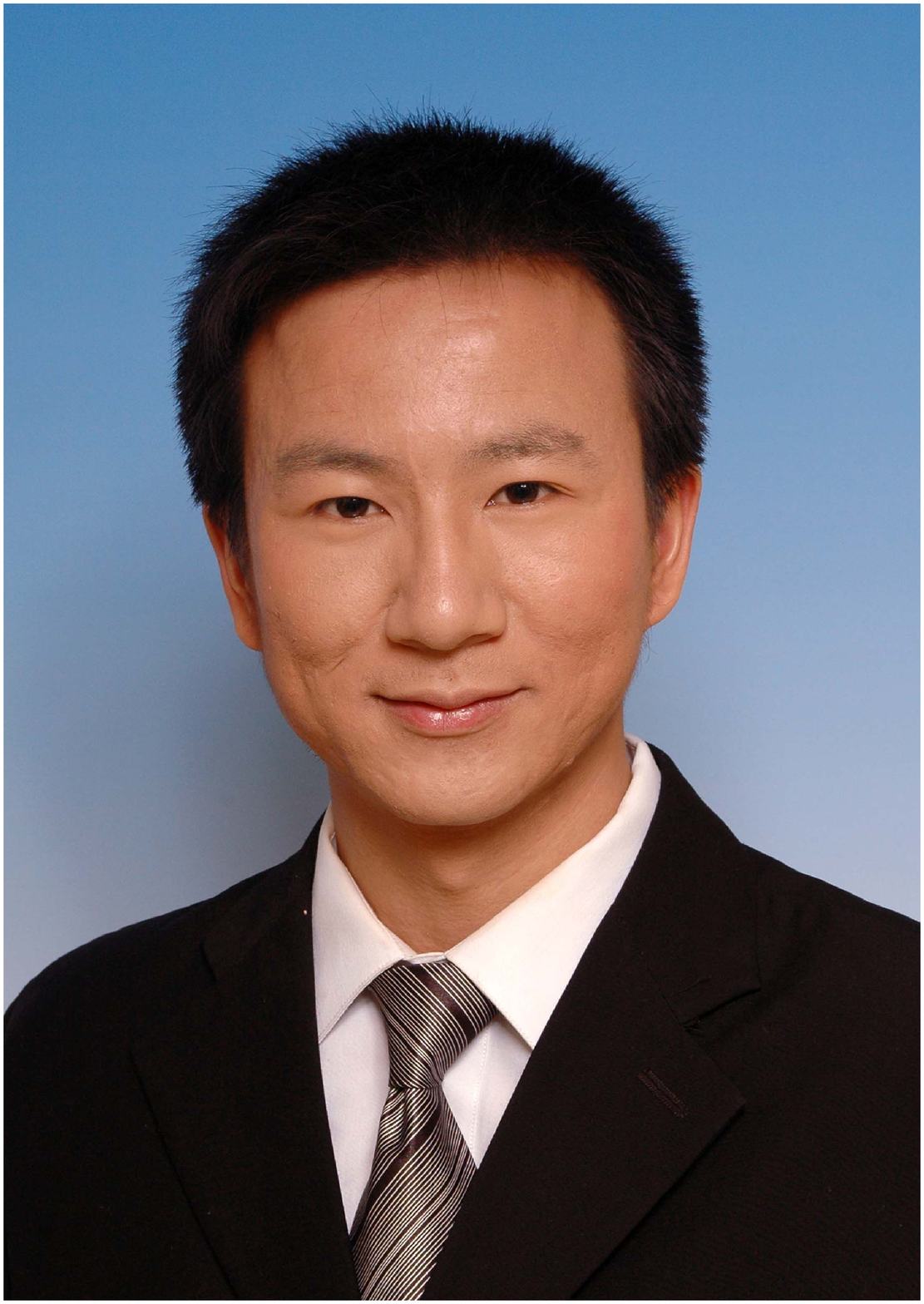}}]{Shaoshi Yang} (S'09) received the B.Eng. Degree in information engineering from Beijing University of Posts and Telecommunications, Beijing, China, in 2006. He is currently working toward the Ph.D. degree in wireless communications with the School of Electronics and Computer Science, University of Southampton, Southampton, U.K., through scholarships from both the University of Southampton and the China Scholarship Council.

From November 2008 to February 2009, he was an Intern Research Fellow with the Communications Technology Laboratory, Intel Labs China, Beijing, where he focused on Channel Quality Indicator Channel design for mobile WiMAX (802.16m). His research interests include multiuser detection/multiple-input mutliple-output detection, multicell joint/distributed processing, cooperative communications, green radio, and interference management. He has published in excess of 20 research papers on IEEE journals and conferences.

Shaoshi is a recipient of the PMC-Sierra Telecommunications Technology Scholarship, and a Junior Member of the Isaac Newton Institute for Mathematical Sciences, Cambridge, UK. He is also a TPC member of both the 23rd Annual IEEE International Symposium on Personal, Indoor and Mobile Radio  Communications (IEEE PIMRC 2012), and of the 48th Annual IEEE International Conference on Communications (IEEE ICC 2013).
\end{IEEEbiography}

\begin{IEEEbiography}
[{\includegraphics[width=1in,height=1.25in,clip,keepaspectratio]{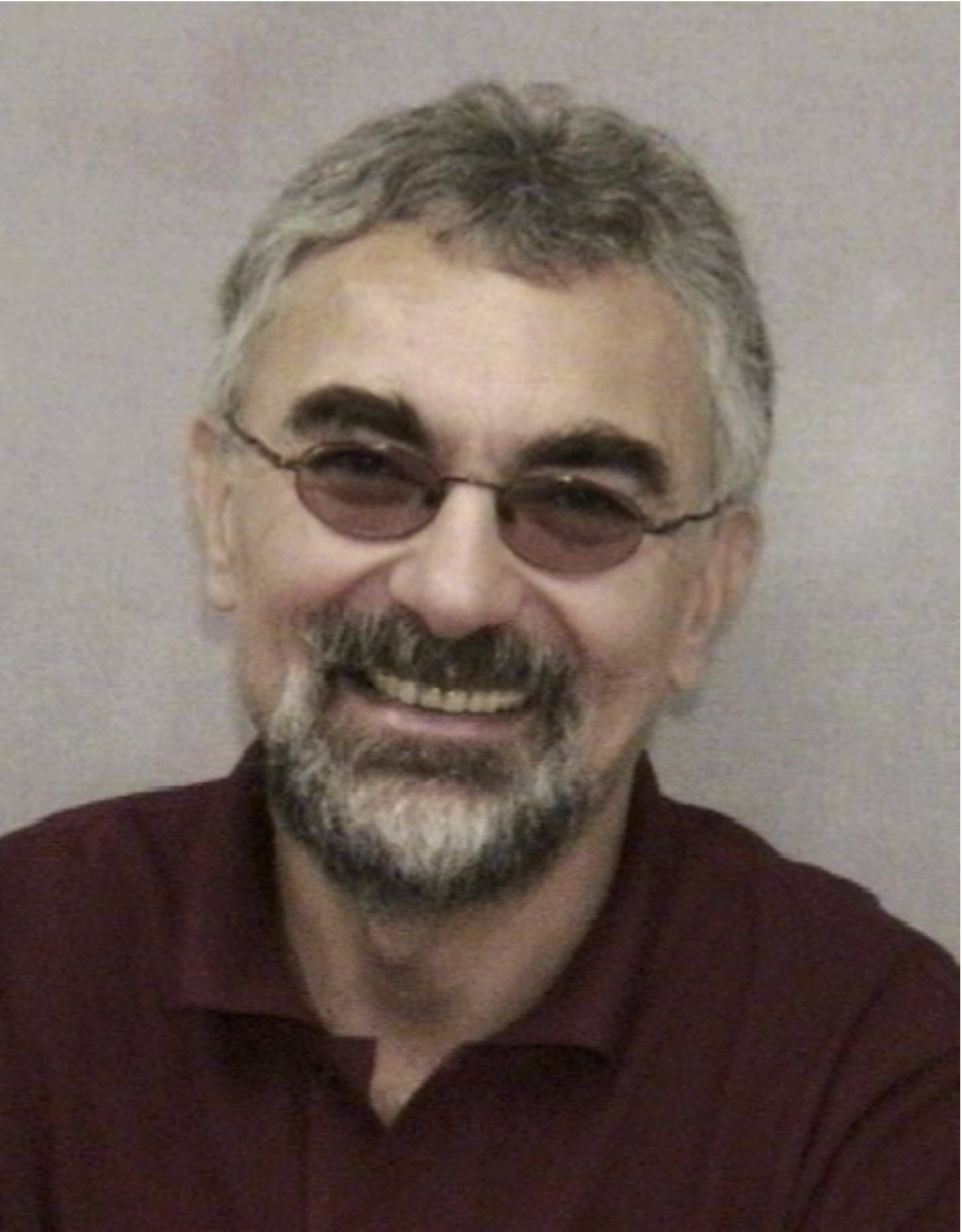}}]{Lajos Hanzo} (http://www.cspc.ecs.soton.ac.uk) FREng, FIEEE, FIET, Fellow of EURASIP, DSc received his degree in electronics in 1976 and his doctorate in 1983. In 2009 he was awarded the honorary doctorate \textit{Doctor Honoris Causa} by the Technical University of Budapest.

During his 35-year career in telecommunications he has held various research and academic posts in Hungary, Germany and the UK. Since 1986 he has been with the School of Electronics and Computer Science, University of Southampton, UK, where he holds the chair in telecommunications. He has successfully supervised 80 PhD students, co-authored 20 John Wiley/IEEE Press books on mobile radio communications totalling in excess of 10000 pages, published 1300+ research entries at IEEE Xplore, acted both as TPC and General Chair of IEEE conferences, presented keynote lectures and has been awarded a number of distinctions. Currently he is directing a 100-strong academic research team, working on a range of research projects in the field of wireless multimedia communications sponsored by industry, the Engineering and Physical Sciences Research Council (EPSRC) UK, the European IST Programme and the Mobile Virtual Centre of Excellence (VCE), UK.

He is an enthusiastic supporter of industrial and academic liaison and he offers a range of industrial courses. He is also a Governor of the IEEE VTS. During 2008-2012 he was the Editor-in-Chief of the IEEE Press and a Chaired Professor also at Tsinghua University, Beijing. His research is funded by the European Research Council's Senior Research Fellow Grant. For further information on research in progress and associated publications please refer to http://www.cspc.ecs.soton.ac.uk Lajos has 17000+ citations.
\end{IEEEbiography} 
\end{document}